\documentclass[aps,prd,superscriptaddress,long,notitlepage,balancelastpage,nofootinbib,floatfix,twocolumn]{revtex4-1}
\pdfoutput=1

\usepackage{amsmath,mathtools,amssymb,amsthm,amsxtra,overpic,bbm,epsfig,subfigure,url,bm}
\usepackage{hyperref}
\usepackage{mathrsfs}
\usepackage{color,xcolor}
\usepackage{comment}
\usepackage{float}
\usepackage{enumitem}
\usepackage{slashed}

\definecolor{nicered}{rgb}{0.5,0.,0.}
\definecolor{nicegreen}{rgb}{0.,0.5,0.}
\definecolor{niceblue}{rgb}{0.,0.,0.5}
\hypersetup{
	colorlinks=true,
	linkcolor=black,
	filecolor=nicegreen,      
	urlcolor=niceblue,
	citecolor=nicered,
}

\newcommand{\prlsection}[2]{{\it\textbf{#1}{#2}}---}
\makeatletter
\newcommand*{\balancecolsandclearpage}{%
	\close@column@grid
	\cleardoublepage
	\twocolumngrid
}
\makeatother

\begin{document}
	\title{ 
		Probing Heavy Sterile Neutrinos at Ultrahigh Energy Neutrino Telescopes  	\\ \vspace{0.05in} 
	via the Dipole Portal 
	}
	\author{Guo-yuan Huang}
	\email{guoyuan.huang@mpi-hd.mpg.de}
	\affiliation{Max-Planck-Institut f{\"u}r Kernphysik, Saupfercheckweg 1, 69117 Heidelberg, Germany} 
	\author{Sudip Jana}
	\email{sudip.jana@mpi-hd.mpg.de}
	\affiliation{Max-Planck-Institut f{\"u}r Kernphysik, Saupfercheckweg 1, 69117 Heidelberg, Germany} 
	\author{Manfred Lindner}
	\email{manfred.lindner@mpi-hd.mpg.de}
	\affiliation{Max-Planck-Institut f{\"u}r Kernphysik, Saupfercheckweg 1, 69117 Heidelberg, Germany} 
	\author{Werner Rodejohann}
	\email{werner.rodejohann@mpi-hd.mpg.de}
	\affiliation{Max-Planck-Institut f{\"u}r Kernphysik, Saupfercheckweg 1, 69117 Heidelberg, Germany} 
	%%%%%%%%%%%%%%%%%%%%%%%%%%
	\begin{abstract}
		\noindent
		The question of how heavy a sterile neutrino can be probed in experiments leads us to investigate the Primakoff production of heavy sterile neutrinos up to PeV masses from ultrahigh-energy neutrinos via the magnetic dipole portal. Despite the suppression from the small magnetic moment, the transition is significantly enhanced by tiny $t$-channel momentum transfers, similar to the resonant production of pions and axions in an external electromagnetic field. Based on the current IceCube measurement of astrophysical neutrinos up to PeV energies, strong constraints can already be derived on the transition magnetic moments of sterile neutrinos up to TeV masses. Moreover, we investigate the sensitivity of future tau neutrino telescopes, which are designed for EeV cosmogenic neutrino detection. We find that sterile neutrino masses as large as $30~{\rm TeV}$ can be probed at tau neutrino telescopes such as GRAND, POEMMA, and Trinity.
	\end{abstract}
	
	\maketitle
	
	\noindent\prlsection{Introduction}{.}%
	The discovery of neutrino flavor oscillations  demonstrates non-zero small neutrino masses and mixings~\cite{Athar:2021xsd,ParticleDataGroup:2020ssz} and the Standard Model (SM) particle spectrum must be enlarged. However, the underlying new physics scale can lie anywhere below the electro-weak (EW) scale up to the Planck scale. If the new physics responsible for the neutrino mass generation mechanism lies at, or slightly above, the EW scale, high-energy collider experiments are able to probe. If new physics lies below the EW scale, neutrino oscillation or scattering experiments can be a better place to look for its direct or indirect implications. However, if the new physics involved lies above the TeV scale, it is less likely to be probed in terrestrial experiments. Here, we are proposing a novel mechanism through which a sterile neutrino of mass up to $\mathcal{O}(30)$ TeV can be probed at  neutrino telescopes, via the transition magnetic moment portal to active neutrinos. 
	
	The study of neutrino magnetic moments began seven decades ago~\cite{Cowan:1954pq}, long before the neutrino was discovered and was a highly popular topic three decades ago owing to the observed sunspot activity~\cite{Davis:1988gd, Davis:1990fb}. Afterwards, several experiments further investigated the neutrino magnetic moments~\cite{Vidyakin:1992nf, Derbin:1993wy, LSND:2001akn, Daraktchieva:2005kn, Deniz:2009mu,  Beda:2012zz, Allen:1992qe, Borexino:2017fbd, Bonet:2022imz}. 
	Theoretical and experimental studies on neutrino magnetic moments became increasingly relevant given that they have the potential to address many unsolved puzzles, including the excess of electron recoil events at XENON1T~\cite{Aprile:2020tmw}, ANITA anomalous events \cite{ANITA:2016vrp,ANITA:2018sgj}, as well as the long-standing MiniBooNE~\cite{MiniBooNE:2018esg} and muon $g-2$ anomalies \cite{Muong-2:2021ojo, Babu:2021jnu}. Triggered by the aforementioned anomalies, the magnetic dipole portal linking the active and sterile neutrinos has been recently received attention and studied at various facilities~\cite{Ismail:2021dyp, Magill:2018jla, Gninenko:2009ks, Schwetz:2020xra, Brdar:2020quo, Shoemaker:2020kji, Bolton:2021pey,  Miranda:2021kre, Vergani:2021tgc,  Zhang:2022spf}. Additionally, active to sterile transition magnetic moments are subject to less severe astrophysical limits than active neutrino magnetic moments~\cite{Raffelt:1999tx, Capozzi:2020cbu, Jana:2022tsa}.
	
	From a theoretical standpoint, the anticipated magnetic moments of neutrinos are imperceptibly tiny in many neutrino mass models that generate the known neutrino masses and mixings \cite{Fujikawa:1980yx, Pal:1981rm}; for a summary, see Ref.~\cite{Giunti:2014ixa}. However, it is conceivable to construct theories consistent with neutrino mass generation that have a quite large neutrino magnetic moments~\cite{Babu:2020ivd}. Thus, understanding the neutrino magnetic moment may give valuable insight into the process by which neutrinos acquire mass and other characteristics.
	
	%%%%%%%%%%%%%%%%%%%%%%%%%%%%%%%%%%%%%%%%%%%%%%%%%%%
	\begin{figure}[t!]
		\begin{center}
			\includegraphics[width=0.45\textwidth]{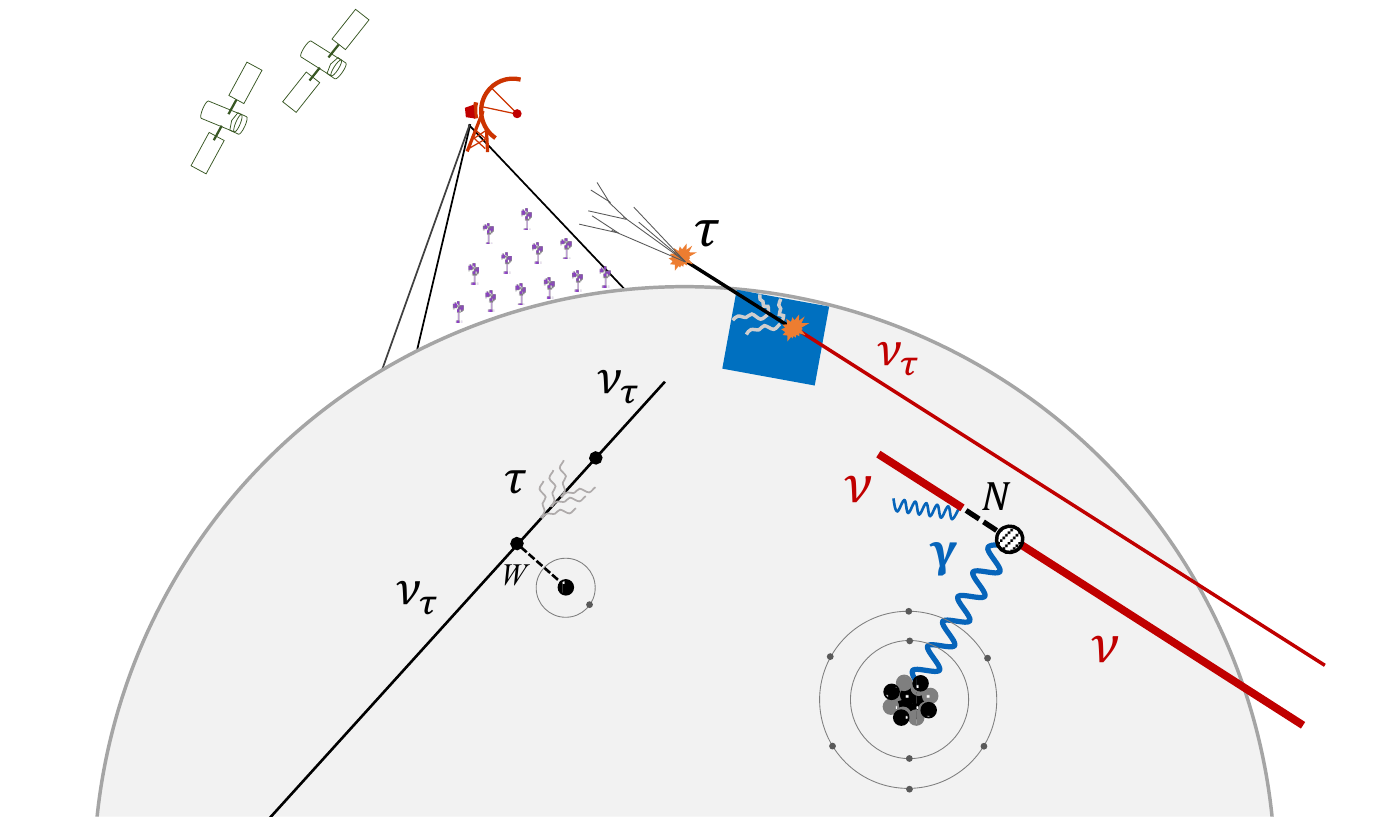}
		\end{center}
		\vspace{-0.2cm}
		\caption{Primakoff production of a heavy right-handed neutrino  from an active neutrino via the magnetic dipole interaction. Due to the long-range nature of the  electromagnetic force, the transition probability features a forward enhancement corresponding to very small momentum transfers.}
		\label{fig:schematic}
	\end{figure}
	%%%%%%%%%%%%%%%%%%%%%%%%%%%%%%%%%%%%%%%%%%%%%%%%%%%
	
	The relevant Lagrangian consisting of a dimension-five operator reads
	\begin{equation} \label{eq:L}
	\mathcal{L} \supset -\mu^{}_{\nu N}\,\overline{\nu^{}_{{\rm L}}} \sigma^{}_{\mu\nu} N F^{\mu\nu},
	\end{equation}
	where $\mu^{}_{\nu N}$ is the transition magnetic moment, $\nu^{}_{\rm L}$ and $N$ denote the active and sterile neutrinos, and $F^{\mu\nu}_{}$ is the usual electromagnetic field strength.
	In fact, such a dipole portal widely exists in many theories with right-handed neutrino extension such as seesaw models. Now, in general, one would anticipate substantial contributions to active neutrino masses in order to get big transition magnetic moments in different ultraviolet extensions of the SM, since both the magnetic moment and mass operators are chirality-flipping.	In terms of the induced Dirac mass term ($m_{\nu N}$), the neutrino transition magnetic moment can be represented as: \begin{equation} \label{eq:mag_scale}
	\frac{\mu_{\nu N}}{\mu_{B}} \sim \frac{2 m_{e} m_{\nu N}}{\Lambda^{2}} \;, 
	\end{equation}
	where $\Lambda$ is the cutoff scale.
	In order to generate neutrino magnetic moments, the loop particles must be electrically charged, and new charged particles below  $ \sim$ 100 GeV \cite{Babu:2019mfe} are typically disfavored from experimental searches. From Eq.~(\ref{eq:mag_scale}), we can see that a neutrino magnetic moment $\mu_{\nu} =10^{-10}\mu_B$ corresponds to a neutrino mass of 1 MeV, which is seven orders of magnitude higher than our current understanding of active neutrino masses. New symmetries and interactions  \cite{Voloshin:1987qy, Babu:1989wn, Babu:2020ivd, Barr:1990um}  may alter the preceding picture, allowing for larger values of $\mu_\nu$ without introducing implausible neutrino masses. For the remainder of our phenomenological study, we choose a framework where the neutrino mass operator, being a Lorentz scalar, is symmetric and forbidden under a new exchange ${\rm SU}(2)_\nu$ symmetry, while the neutrino magnetic moment operator,  being a Lorentz tensor, is anti-symmetric and allowed under the ${\rm SU}(2)_\nu$ symmetry \cite{Voloshin:1987qy} and will remain agnostic about the possible connection between the magnetic moment and the neutrino masses.
	In order to have a consistent effective field theory description of $\nu N$ interactions as in Eq.~(\ref{eq:L}), particles inside the loop should carry masses larger than the sterile neutrino mass $m^{}_{N}$ and the scattering energy scale. For lighter loop particles, the effective operator is not sufficient to capture the right dynamics, and we should consider loop particles as active degrees of freedom which can be directly produced. We concentrate on the effective magnetic dipole portal in the present work. The latter case is very model-dependent, and we shall leave it for future studies. 
	
	%%%%%%%%%%%%%%%%%%%%%%%%%%%%%%%%%%%%%%%%%%%%%%%%%%%
	\begin{figure}[t!]
		\begin{center}
			\includegraphics[width=0.45\textwidth]{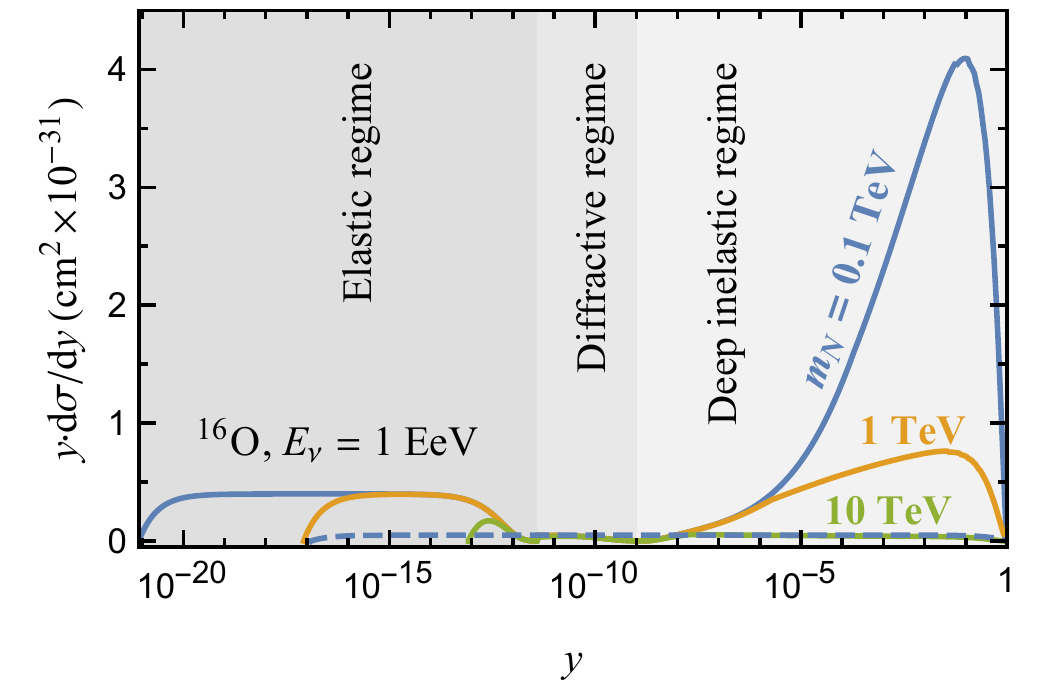}
		\end{center}
		\vspace{-0.3cm}
		\caption{The differential cross section of $\nu \to N$ conversion with respect to the energy loss parameter $y$ assuming an ${}^{16}{\rm O}$ target. The energy loss of the conversion is parameterized by $y E^{}_{\nu}$ with $E^{}_{\nu}=1~{\rm EeV}$. 
			The magnetic moment is chosen as $\mu^{}_{\nu N} = 10^{-4}~{\mu}^{}_{\rm B}$.
			The cases with $m^{}_{N} = 0.1~{\rm TeV}$, $1~{\rm TeV}$ and $10~{\rm TeV}$ are shown as blue, yellow and green curves, respectively.	
			For $m^{}_{N} = 0.1~{\rm TeV}$, three regimes including the elastic, diffractive and deep inelastic scatterings are indicated by the gray shaded regions. The subdominant neutrino-electron scattering is given as the blue dashed curve (averaged over nucleon number) for comparison.}
		\label{fig:xsec_y}
	\end{figure}
	%%%%%%%%%%%%%%%%%%%%%%%%%%%%%%%%%%%%%%%%%%%%%%%%%%%
	A notable consequence of  sizable magnetic moments is that neutrinos will have long-range interactions besides the short-range one suppressed by heavy $W$ and $Z$ masses.
	The $t$-channel exchange of a massless photon will result in a significant enhancement of cross section in the forward direction.
	The enhancement of $N$ production is analogous to the Primakoff effect~\cite{Primakoff:1951iae}, which describes the resonant production of neutral scalars such as pions and axions in an external electromagnetic field.
	To be more specific, the cross section for the transition $\nu \to N$ via the dipole portal will be inversely proportional to the momentum transfer square $Q^2$, namely $\mathrm{d}\sigma/\mathrm{d} Q^2 \propto 1/Q^2$.
	
	The minimum momentum transfer required by such a transition is determined by the incoming neutrino energy and the sterile neutrino mass, i.e., $Q^2_{\rm min} \approx m^4_N/(4 E^2_{\nu}) $,  which corresponds to an impact parameter $b \sim 2E^{}_{\nu}/m^{2}_{N}$.
	The result can be understood from the fact that with higher and higher neutrino energies, it will become easier and easier to put the highly boosted $N$ on-shell for the same tiny moment transfer.
	Consequently, we are able to probe larger sterile neutrino masses $m^{}_{N}$ by increasing the incoming neutrino energy $E^{}_{\nu}$.
	
	To date, the highest detected neutrino energy is several PeV by the IceCube Observatory embedded in Antarctic ice~\cite{IceCube:2013cdw}.
	The energy frontier of ultrahigh energy (UHE) neutrinos can be further pushed by the future projects like Ashra-NTA~\cite{Ogawa:2021dK}, BEACON~\cite{Wissel:2020fav,Wissel:2020sec}, GRAND~\cite{GRAND:2018iaj,Kotera:2021ca}, IC-Gen2 Radio~\cite{IceCube-Gen2:2020qha,Hallmann:2021kqk}, POEMMA~\cite{POEMMA:2020ykm}, TAMBO~\cite{Romero-Wolf:2020pzh}, Trinity~\cite{Otte:2018uxj,Otte:2019aaf,Wang:2021/M,Brown:2021tf} and so on (see Refs.~\cite{Huang:2021mki, Abraham:2022jse,Ackermann:2022rqc, Arguelles:2022xxa} for recent reviews), which are designed for the detection of cosmogenic neutrinos~\cite{Berezinsky:1969qj,Greisen:1966jv,Zatsepin:1966jv} with energies as high as $1~{\rm EeV} \equiv 10^{9}~{\rm GeV}$.
	Various phenomenological potentials of this energy frontier have been investigated; for an incomplete list of references, see a partial list of Refs.~\cite{Jezo:2014kla,Jho:2018dvt,Huang:2019hgs,Denton:2020jft,Valera:2021dix,Soto:2021vdc,Valera:2022ylt,Xing:2011zm,Babu:2019vff,Bustamante:2020niz,Zhou:2020oym,Song:2020nfh,Dey:2020fbx, Dev:2016uxj, Babu:2022fje}.

	A promising class of cosmogenic neutrino observatories is the tau neutrino telescope~\cite{Berezinsky:1975zz,Domokos:1997ve,Domokos:1998hz,Capelle:1998zz,Fargion:1999se,Fargion:2000iz,LetessierSelvon:2000kk,Feng:2001ue,Kusenko:2001gj,Bertou:2001vm,Cao:2004sd,Baret:2011zz}, which can achieve an unprecedented effective area by using high-altitude detectors to monitor the $\nu^{}_{\tau}$-induced extensive air shower from a mountain or the Earth surface.
	Given the successful UHE neutrino observation~\cite{IceCube:2013low,IceCube:2013cdw,IceCube:2018cha,IceCube:2021rpz,IceCube:2020abv,ANTARES:2011hfw} and an ongoing exciting campaign of EeV cosmogenic neutrino detection~\cite{Adams:2017fjh,Eser:2021H6,Cummings:2020ycz,Abarr:2020bjd,Vieregg:2021nC,POEMMA:2020ykm,GRAND:2018iaj,Kotera:2021ca,Romero-Wolf:2020pzh,Ogawa:2021dK,Otte:2018uxj,Otte:2019aaf,Wang:2021/M,Brown:2021tf,Wissel:2020fav,Wissel:2020sec,IceCube-Gen2:2020qha,Hallmann:2021kqk,RNO-G:2020rmc,Aguilar:2021uzt,ARA:2019wcf,Anker:2020lre,deVries:2021BA,Prohira:2021vvn,Prohira:2019glh}, it is very timely to investigate the possibility of heavy sterile neutrino production  at those facilities.	
	A schematic diagram with the relevant Primakoff conversion process is illustrated in Fig.~\ref{fig:schematic}, where the incoming neutrino flux can be severely affected by the conversion process before reaching the neutrino detector.
	The UHE neutrinos are then  detected by an in-ice volume (for all neutrino flavors), and an atmospheric radio or imaging telescope (for tau neutrinos).
	
	\prlsection{Conversion cross section from TeV to EeV energies}{.}%
	%Analytical discussion.  Why the cross section is large at extremely high energies.
	%
	%%%%%%%%%%%%%%%%%%%%%%%%%%%%%%%%%%%%%%%%%%%%%%%%%%%
	\begin{figure}[t]
		\begin{center}
			\includegraphics[width=0.45\textwidth]{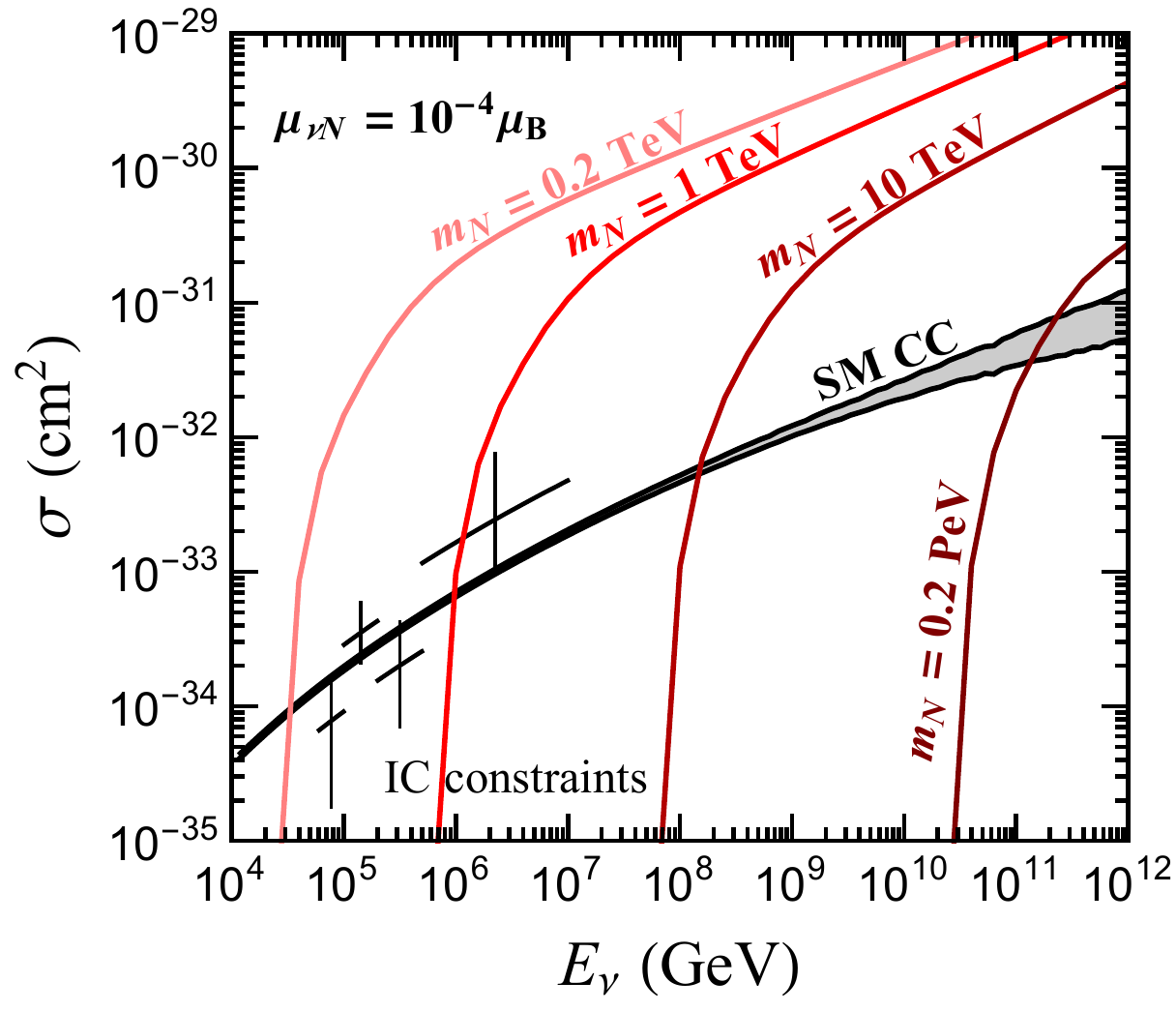}
		\end{center}
		\vspace{-0.3cm}
		\caption{The neutrino cross sections of conversion to a sterile neutrino with mass $m^{}_{N}$ while propagating in ice or water ${\rm H}^{}_{2}{\rm O}$. The cross section has been averaged over nucleons. The transition magnetic moment $\mu^{}_{\nu N}$ is fixed to $10^{-4} \mu^{}_{\rm B}$. For comparison, the neutrino charged-current cross section in the SM is also shown. Choosing a different value of $\mu^{}_{\nu N}$ will shift the whole cross section curve upwards or downwards. The IceCube constraint on the neutrino cross section below 10~PeV is shown for comparison.}
		\label{fig:xsec_Enu}
	\end{figure}
	%%%%%%%%%%%%%%%%%%%%%%%%%%%%%%%%%%%%%%%%%%%%%%%%%%%
	Different from the SM CC interaction, the $t$-channel exchange of a massless photon features a significant forward enhancement, which allows us to probe small electromagnetic couplings.
	Depending on the magnitude of $Q^2$, we may run into three different regimes~\cite{Magill:2018jla}: (i) the elastic coherent scattering off the whole nucleus; (ii) the diffractive (quasi-elastic) scattering off each individual nucleons with $A$ being the mass number of the target atom; (iii) the deep inelastic scattering (DIS) with partons in the nucleon.
	For SM processes, the deep inelastic scattering dominates the reaction of UHE neutrinos. Interestingly, the $\nu$-$N$ conversion  of our concern gains from small momentum transfers, such that the coherent scattering in some cases can be dominant.
	The technical details of calculating the cross section for different regimes are given in the Appendix.

	Fig.~\ref{fig:xsec_y} shows the differential cross section as a function of the energy loss parameter $y$, with the incoming neutrino energy $E^{}_{\nu} = 1~{\rm EeV}$ and ${}^{16} {\rm O}$ as the target. The actual energy loss and momentum transfer $Q^2$ are given by $y E^{}_{\nu}$ and $2 M E^{}_{\nu} y$, respectively, where $M$ is the target mass (nucleus, nucleon or parton). 
	%The total cross section can be read from the area in Fig.~\ref{fig:xsec_y}.
	%
	The blue, yellow and green curves correspond to $m^{}_{N} = 0.1~{\rm TeV}$, $1~{\rm TeV}$ and $10~{\rm TeV}$, respectively.
	The parameter spaces of $y$ corresponding to $Q^2$ of the three different regimes mentioned above are indicated by the shaded regions.
	As has been mentioned, $Q^2$ is bounded from below by $Q^2_{\rm min} \approx m^4_N/(4 E^2_{\nu})$.
	At extremely small $y$ or $Q^2$, neutrinos interact collectively with the whole nucleus, and the differential cross section $y \cdot {\rm d}\sigma/{\rm d} y$ is almost a constant.
	At larger $y$, the quasi-inelastic diffractive reaction becomes efficient. For even larger $y$, the exchanged photon probes individual partons and the DIS comes into play.
	The DIS contribution increases with smaller $m^{}_{N}$, because there will be more sea quarks contributing at small Bjorken-$x$.

\begin{figure*}[t!]
		\begin{center}
			\includegraphics[width=0.4\textwidth]{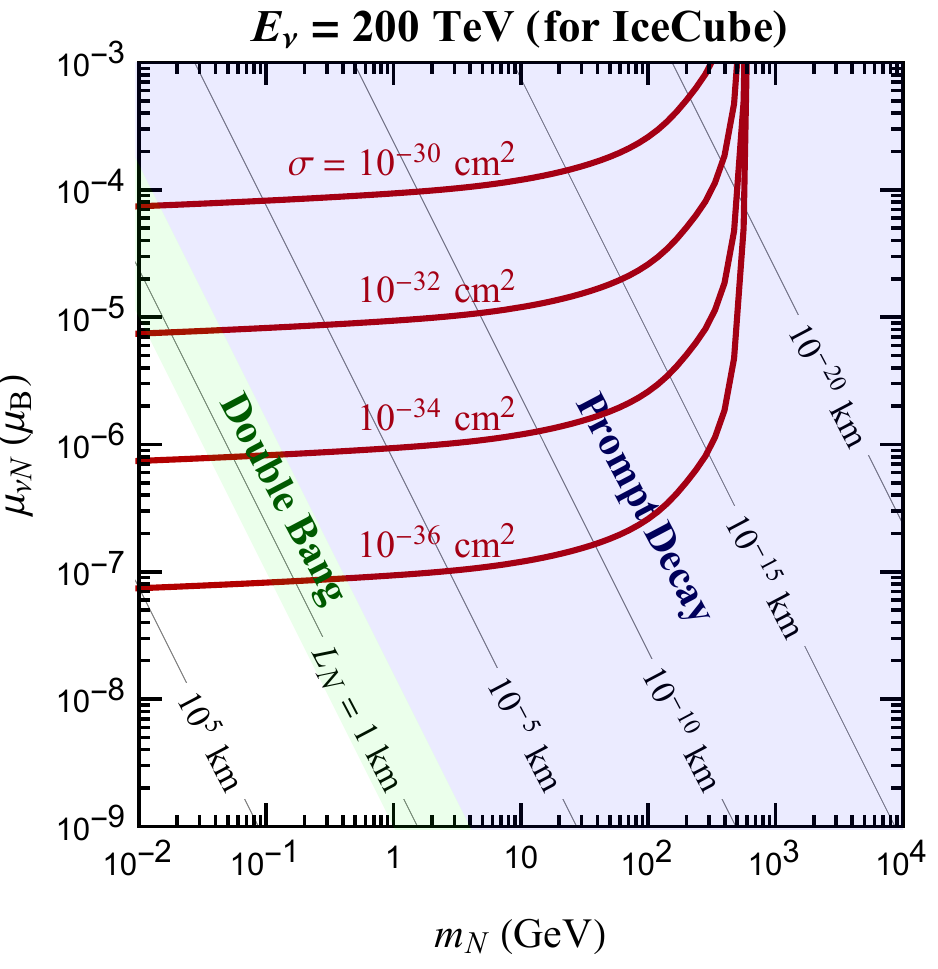}
			\includegraphics[width=0.4\textwidth]{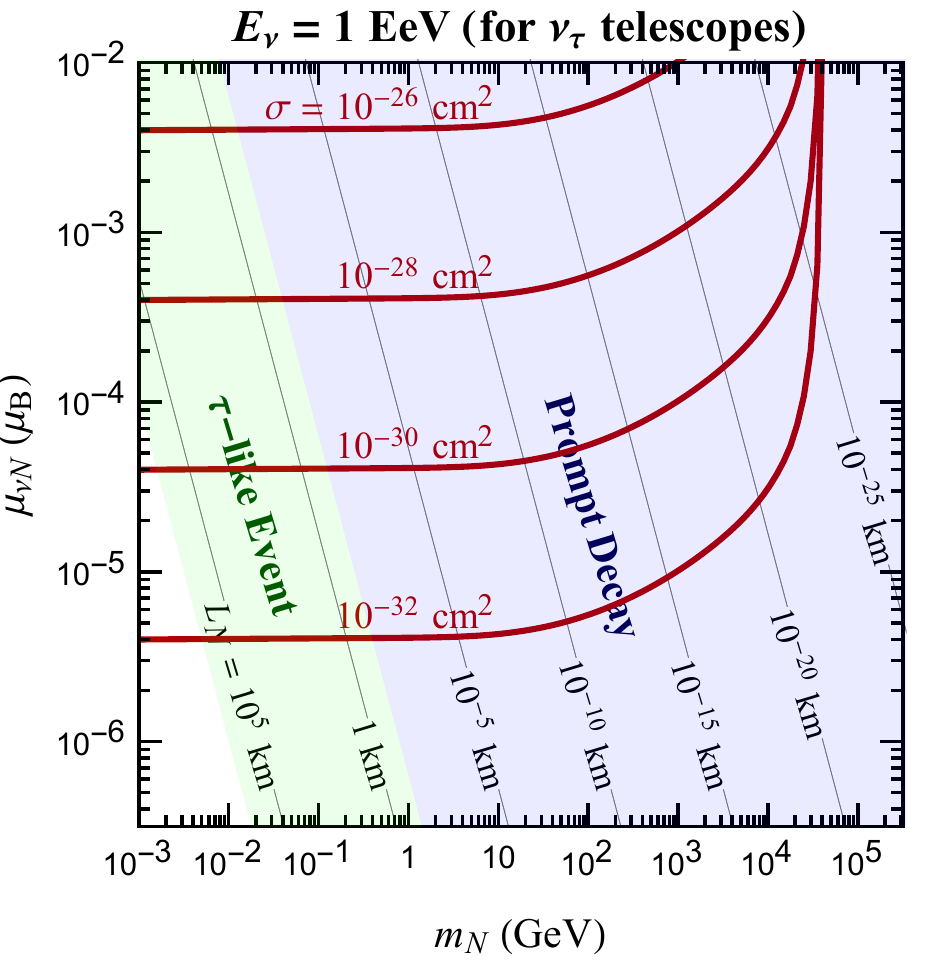}
		\end{center}
		\vspace{-0.3cm}
		\caption{Contours of nucleon-averaged cross sections (red curves) in the parameter space of sterile neutrino mass $m^{}_{N}$ and transition magnetic moment $\mu^{}_{\nu N}$. The collision target here is fixed as ${\rm H}^{}_{2}{\rm O}$ for demonstration. The left and right panels stand for the cases with energies $E^{}_{\nu} = 200~{\rm TeV}$ and $E^{}_{\nu} = 1~{\rm EeV}$, relevant for IceCube and tau neutrino telescopes, respectively. 
			The gray lines represent the decay length of the sterile neutrino in the corresponding parameter space.
		}
		\label{fig:xsec_contour}
	\end{figure*}

	At high energy scales, the neutrino  cross section will benefit from the sea quark enhancement.
	In Fig.~\ref{fig:xsec_Enu} we compare the cross section of $\nu$-$N$ conversion to that of the charged-current $\nu$-$\ell$ conversion in the SM \cite{Gandhi:1998ri}, with respect to the incoming neutrino energy $E^{}_{\nu}$.
	From lighter to darker red colors, we illustrate the cases of $m^{}_{N} = 0.2~{\rm TeV}$, $1~{\rm TeV}$, $10~{\rm TeV}$ and $0.2~{\rm PeV}$, respectively. The SM cross section with the uncertainty of parton distribution functions is given as the gray shaded band.
	Note that for each case, there is a lower cutoff of the cross section, which corresponds to the energy threshold to make $N$ accessible, namely $E^{\rm min}_{\nu} \approx m^2_{N}/(2M)$.
	Well above this threshold, the $\nu$-$N$ conversion exhibits a similar power law relation as the SM CC process, with respect to neutrino energy. This is due to that the DIS becomes the dominant contribution at higher neutrino energies.
	The transition magnetic moment in Fig.~\ref{fig:xsec_Enu} has been fixed to $\mu^{}_{\nu N} = 10^{-4}\mu^{}_{\rm B}$, and the cross section can be simply rescaled if a different $\mu^{}_{\nu N}$ has been adopted according to $\sigma \propto \mu^{2}_{\nu N}$. 
	%Numerical discussion, comment on Figs 3 and 4.
	To illustrate the distributions of cross sections in the parameter space of our concern, in Fig.~\ref{fig:xsec_contour} we show the contours of cross sections, as red curves, in the $m^{}_{N}$-$\mu^{}_{\nu N}$ plane. For the left (right) panel, the neutrino energy has been taken to be $E^{}_{\nu} = 200~{\rm TeV}$ ($E^{}_{\nu} = 1~{\rm EeV}$), which is more relevant for IceCube (tau neutrino telescopes).
	After production, $N$ decays to $\nu \gamma$ via the dipole portal.
	The decay length of the produced $N$ reads, $L^{}_{N} = 4\pi E^{}_{\nu}/( \mu^{2}_{\nu N} m^4_{N})$, which has been indicated by the gray lines.
	Depending on the decay length, there can be two different scenarios: (i) $N$ decays promptly inside the target, and hence will increase the attenuation effect of neutrinos in matter; (ii) $N$ is long-lived and can be separately identified by the detector.

	For the parameter space of our interest with very large $m^{}_{N}$, $N$ decays promptly after production and affects the neutrino interaction in matter. Nevertheless, let us comment on the scenario where $N$ is light enough to be long-lived.
	For IceCube, the sterile neutrino decay will possibly induce a double bang event inside the Cherenkov volume, if the decay length is shorter than the volume size ($\sim 1~{\rm km}$) but longer than the resolvable distance. 
	Similarly, at tau neutrino telescopes designed for tau detection, the long-lived $N$ will induce tau-like events if the decay length is comparable or longer than that of tau, $L^{}_{\tau} \approx 50~{\rm km}\cdot (E^{}_{\nu}/{\rm EeV})$.
	Such tau-like events can be well distinguished from SM tau decays in the case that $N$ is long-lived enough, and  extensive air showers  coming from very steep angles similar to the anomalous ANITA events will be expected.
	%
	%For mountain telescopes such as GRAND and Trinity, the decay of $N$ will be detectable as long as the decay length is comparable to the mountain size, which can be as thin as $1~{\rm km}$.
	%
	However, the requirement of the decay length will limit $m^{}_{N}$ to be smaller than $1~{\rm GeV}$. 
	For small sterile neutrino masses, the transition magnetic moment receives severe constraints from DONUT~\cite{DONUT:2001zvi}, XENON1T~\cite{Brdar:2020quo}, Big Bang nucleosynthesis~\cite{Brdar:2020quo} and Supernova 1987A~\cite{Magill:2018jla}, and can be significantly improved with future probes like atmospheric neutrino searches~\cite{Coloma:2017ppo,Atkinson:2021rnp}, SHiP~\cite{Magill:2018jla,Alekhin:2015byh}, DUNE~\cite{Schwetz:2020xra} and the forward facility of LHC~\cite{Ismail:2021dyp}. 
	Above $m^{}_{N} = 10~{\rm GeV}$, only collider experiments like LEP and LHC are able to put competitive constraints~\cite{Magill:2018jla}. However, due to the limitation of the center-of-mass (COM) energy $\sqrt{s}$, the LEP limit will stop around $m^{}_{N}=100~{\rm GeV}$, and LHC will lose sensitivity above $m^{}_{N} = 1~{\rm TeV}$ let alone the large backgrounds of mono-photon searches. 
	
	\prlsection{Constraints and sensitivities}{.}%
	UHE neutrino measurements can reach much higher COM energies, e.g., $\sqrt{s} = 1.4~{\rm TeV}$ with $E^{}_{\nu} = 1~{\rm PeV}$ and $\sqrt{s} = 0.14~{\rm PeV}$ with $E^{}_{\nu} = 10~{\rm EeV}$ for neutrino-nucleon scatterings.
	Let us proceed to explore the power of the currently running UHE neutrino observatory IceCube as well as the sensitivity of future telescopes, for which we will take the mountain-valley telescope GRAND~\cite{GRAND:2018iaj,Kotera:2021ca}, the satellite telescope POEMMA~\cite{POEMMA:2020ykm} and the mountain-top telescope Trinity~\cite{Otte:2018uxj,Otte:2019aaf,Wang:2021/M,Brown:2021tf} as three representative examples.

	%Short intro to IceCube
	After data taking since as early as 2005, the IceCube Observatory has clearly detected an UHE neutrino flux of  astrophysical origin.
	By analyzing the angular distribution of events, the IceCube data have been used to set constraints on neutrino scattering cross sections up to PeV energies~\cite{Bustamante:2017xuy,IceCube:2017roe,IceCube:2020rnc}, which are recast as the $1\sigma$ error bars in Fig.~\ref{fig:xsec_Enu}.

\begin{figure}[b!]
		\begin{center}
			\includegraphics[width=0.4\textwidth]{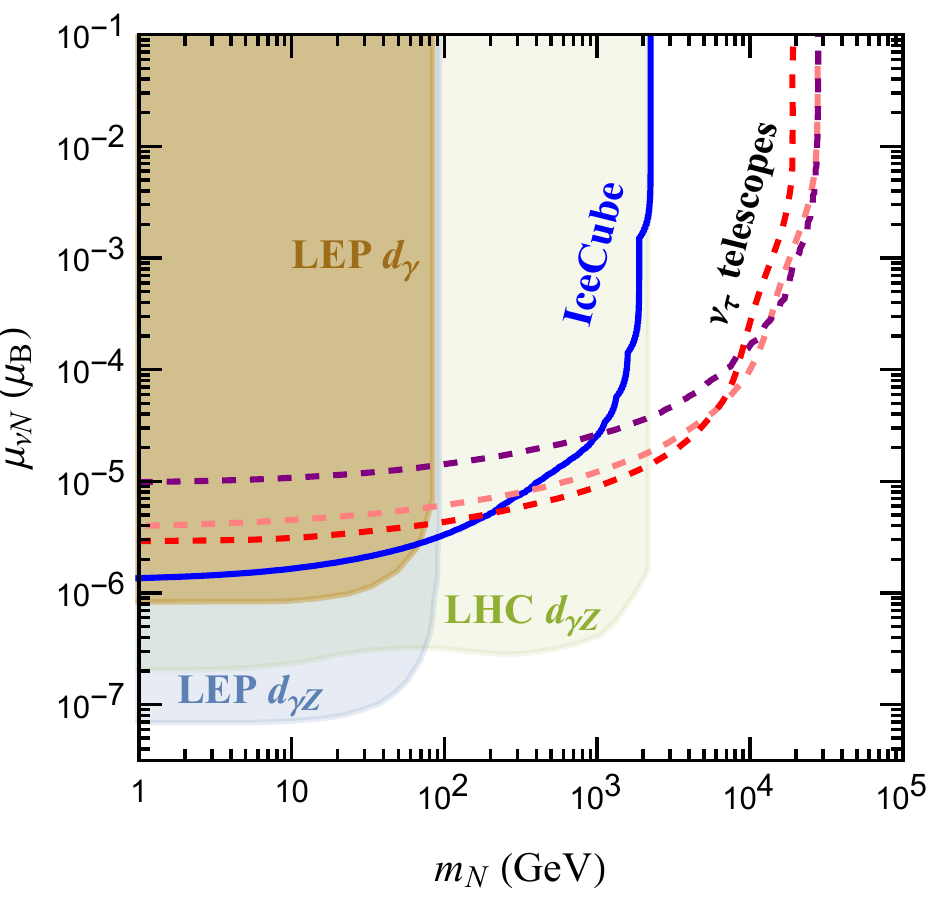}
		\end{center}
		\vspace{-0.3cm}
		\caption{The $3\sigma$ exclusion and sensitivity contours of IceCube and tau neutrino telescopes (purple for GRAND, red for POEMMA, and pink for Trinity) to the neutrino transition magnetic moment. Note that for the IceCube curve, we assume that the three neutrino flavors share the same transition dipole magnetic moment. For tau neutrino telescopes, only the least constrained $\mu^{}_{\nu_{\tau} N}$ participates. }
		\label{fig:sens}
	\end{figure}

	The constraints on neutrino transition magnetic moments can be  obtained by properly translating from the existing IceCube measurement, using our cross section calculation in terms of $m^{}_{N}$ and $\mu^{}_{\nu N}$.
	The chi-square $\chi^2_{\rm IC}$ can be easily constructed by using the central and $1\sigma$ values fitted by IceCube; see the Appendix for more details.
	Taking $\chi^2_{\rm IC} = 11.83$, the IceCube $3\sigma$ constraint on the sterile neutrino is shown as the blue curve in Fig.~\ref{fig:sens}. 
	%
	%Some comments on previous discussion of atmospheric neutrinos
	Notice that previous attempts to use the atmospheric neutrino detection at IceCube/DeepCore~\cite{Coloma:2017ppo} and Super-Kamiokande~\cite{Atkinson:2021rnp} with double bang events can set strong constraints on neutrino transition magnetic moment, though only for very small $m^{}_{N}$ (below $1~{\rm GeV}$) due to the low energies considered.
	%Set constraints based on IceCube cross section measurement.

	%Tau nu tel introduction
	To reach even larger $N$ masses, the detection of a neutrino flux with higher energies is required.
	There is such a guaranteed flux, namely cosmogenic neutrinos with energy as high as EeV, generated from cosmic rays colliding with the cosmic microwave and cosmic infrared backgrounds during propagation. 
	Those cosmogenic neutrinos can be efficiently detected by a large number of future projects.

	The event analysis of tau neutrino telescopes typically requires to simulate the neutrino and tau propagation. Both the tau decay and energy loss should be properly taken into account for the simulation.
	The presence of neutrino transition magnetic moment will modify the neutrino propagation:
	\begin{widetext}
		\begin{align}   \label{eq:nupropa1}
		\frac{\mathrm{d}}{\mathrm{d} t} \left(\frac{\mathrm{d}\Phi^{}_{\nu} }{ \mathrm{d} E^{}_{\nu}}  \right)  = & -\; N^{}_{\rm A} \rho \left(\sigma^{\rm CC}_{\rm SM} + \sigma^{\rm NC}_{\rm SM} +  {\sigma^{\rm }_{ \nu N}} \right)  \frac{\mathrm{d}\Phi^{}_{\nu} }{ \mathrm{d} E^{}_{\nu}}  + \;  N^{}_{\rm A} \rho \int \mathrm{d} E^{\prime}_{\nu}
		\frac{\mathrm{d}\Phi^{}_{\nu} }{ \mathrm{d} E^{\prime}_{\nu}}  \frac{1}{E^{\prime}_{\nu} } \left.\left(\frac{\mathrm{d}\sigma^{\rm NC}_{\rm SM}}{\mathrm{d} z}+ \frac{\mathrm{d} {\sigma^{\rm }_{ \nu N}}}{\mathrm{d} z} \right)\right|_{z = \frac{E^{}_{\nu}}{E^{\prime}_{\nu}} } \notag\\
		& + \int \mathrm{d} E^{\prime}_{ \tau} \frac{\mathrm{d} \Phi^{}_{\tau}}{\mathrm{d} E^{\prime}_{ \tau}} 
		\frac{1}{E^{\prime}_{\tau}} \frac{\mathrm{d} \Gamma^{}_{\tau} }{ \Gamma^{}_{\tau} \mathrm{d} z} \;,
		\\ 
		\label{eq:nupropa2}
		\frac{\mathrm{d}}{\mathrm{d} t} \left(\frac{\mathrm{d}\Phi^{}_{\tau} }{ \mathrm{d} E^{}_{\tau}}  \right)  = & - \; \Gamma^{}_{\tau} \frac{\mathrm{d}\Phi^{}_{\tau} }{ \mathrm{d} E^{}_{\tau}} 
		- N^{}_{\rm A} \frac{\rho}{A} \sigma^{}_{\rm \tau}\frac{\mathrm{d}\Phi^{}_{\tau} }{ \mathrm{d} E^{}_{\tau}}   
		+ \;  N^{}_{\rm A} \frac{\rho}{A}\int \mathrm{d} E^{\prime}_{\tau} \frac{\mathrm{d}\Phi^{}_{\tau}}{\mathrm{d} E^{\prime}_{\tau}} \frac{1}{E^{\prime}_{\tau}} 
		\left. \frac{\mathrm{d}\sigma^{}_{ \tau}  }{\mathrm{d} z} \right|_{z = \frac{E^{}_{\tau}}{E^{\prime}_{\tau}} }
		+ \; \rho \frac{\partial}{\partial E^{}_{\tau}} \left(  \beta^{}_{\tau} E^{}_{\tau} \frac{\mathrm{d}{\Phi^{}_{\tau}}}{\mathrm{d} E^{}_{\tau}}   \right) 
		\notag\\
		&  
		+ \; N^{}_{\rm A} \rho \int \mathrm{d} E^{\prime}_{\nu} \frac{\mathrm{d} \Phi^{}_{\nu}}{\mathrm{d} E^{\prime}_{\nu}} \frac{1}{ E^{\prime}_{ \nu }} \left. \frac{\mathrm{d}\sigma^{\rm CC}_{\rm SM}}{\mathrm{d} z} \right|_{z = \frac{E^{}_{\tau}}{E^{\prime}_{\nu}} } \;.
		\end{align}
	\end{widetext}
	Here $\sigma^{}_{\nu N}$ is the new-physics nucleon-averaged cross section of $\nu$-$N$ conversion, for which the target nucleus should be specified during the propagation, e.g., ${\rm H}^{}_{2} {\rm O}$ for water and ice, ${\rm Si}{\rm O}^{}_{2}$ for rock.
	Note that the immediate decay of $N$ into $\nu \gamma$ will regenerate the neutrino flux.
	The fluxes of neutrinos and taus are represented by
	$\mathrm{d}\Phi^{}_{\nu} / \mathrm{d} E^{}_{\nu}$ and $\mathrm{d}\Phi^{}_{\tau} / \mathrm{d} E^{}_{\tau}$, respectively, $N^{}_{\rm A}$ is the Avogadro constant, $\rho$ is the matter density, $A$ is the mass number of the target nucleus, and $z \equiv 1-y$.
	
	%\rd{event distribution demo.}
		%%%%%%%%%%%%%%%%%%%%%%%%%%%%%%%%%%%%%%%%%%%%%%%%%%%
	%%%%%%%%%%%%%%%%%%%%%%%%%%%%%%%%%%%%%%%%%%%%%%%%%%%
	%	
	%%%%%%%%%%%%%%%%%%%%%%%%%%%%%%%%%%%%%%%%%%%%%%%%%%%
	%%%%%%%%%%%%%%%%%%%%%%%%%%%%%%%%%%%%%%%%%%%%%%%%%%%
	
	With the above simulated tau flux, the event integration can be performed for given experimental configurations. We follow the proposed setups of GRAND (200k radio antennas and ten-year running), POEMMA (in Limb mode, $360^{\circ}$ field of view and five-year running) and Trinity (three stations and ten-year running) as closely as possible; see Appendix and Ref.~\cite{Huang:2021mki} for more technical details. The GRAND antenna screen is assumed to be placed on a mountain slope inclined by $3^{\circ}$, which will probably change for the final locations but will not influence our results significantly.
	We then generate the events at tau neutrino telescopes for each choice of $m^{}_{N}$ and $\mu^{}_{\nu N}$ in our parameter space, and calculate the sensitivity based on the chi-square
	\begin{eqnarray}
	\chi^2_{\rm min}= \underset{\Phi^{\rm in}_{\nu}}{\rm Min} \left\{ \sum_{i = 1}^{N_{\rm bins}} \frac{\left( n_{i}^{\rm th} - n_{i}^{\rm exp} \right)^2}{n_{i}^{\rm th} } \right\},
	\end{eqnarray}
	where $n^{\rm exp(th)}_{i}$ is the experimental (theoretical) event number in the $i$-th bin of the energy and angular space. 
	The experimental events are produced by assuming a cosmogenic neutrino flux~\cite{Murase:2015ndr} and null new physics contributions, which is to be compared to the theory predictions with any given $m^{}_{N}$ and $\mu^{}_{\nu N}$. The chi-square is then minimized over the incoming neutrino flux $\Phi^{\rm in}_{\nu}$ when we calculate the theory predictions.
	%Description of the telescopes

	The obtained $3\sigma$ sensitivities are shown as the purple (GRAND), red (POEMMA) and pink (Trinity) curves in Fig.~\ref{fig:sens}.
	Collider experiments including LEP and LHC have put limits on the transition magnetic moment above $m^{}_{N} = 1~{\rm GeV}$ based on  mono-photon searches~\cite{Magill:2018jla}, which are shown as the blue, yellow and green shaded regions in Fig.~\ref{fig:sens}.
	The LHC limit (in green) translated from dark matter searches is set on the dipole coupling to the combination of $\gamma$ and $Z$ bosons, which is hence more aggressive than the minimal scenario with only photon coupling. 
	Nevertheless, the proposed tau neutrino telescopes such as GRAND, POEMMA and Trinity are able to achieve unprecedented sensitivity to the sterile neutrino mass, i.e., as high as $30~{\rm TeV}$, via the dipole portal.

	\prlsection{Final remarks}{.}%
	UHE neutrinos are not only a perfect messenger for astronomy, but also a promising probe of new physics beyond the Standard Model.
	The EeV cosmogenic neutrino flux, though uncontrollable, represents a new energy frontier with collision energies much higher than that has been achieved by colliders.
	With this energy frontier, UHE neutrino telescopes have many advantages in probing certain new physics processes.
	We have shown that with a small dipole coupling, the conversion of active neutrinos to sterile neutrinos as heavy as $10~{\rm TeV}$ can become efficient when the neutrino energy reaches $1~{\rm EeV}$.
	The future neutrino telescopes aiming at cosmogenic neutrino detection, such as GRAND, POEMMA and Trinity, are able to push the sterile neutrino mass to $30~{\rm TeV}$ if a large transition magnetic moment is present.
	
	Even though we have focused on large $m^{}_{N}$ in the present work, we comment on another possible effect of small $m^{}_{N}$ using the UHE neutrino observations.
	Through the dipole portal, the collision of UHE neutrinos with the cosmic photon background, e.g. $\nu+\gamma^{}_{\rm CMB} \to N$, may lead to dip structures in the UHE neutrino spectrum, similar to the previous discussion on UHE neutrino scattering off C$\nu$B in the context of secret neutrino interactions~\cite{Bustamante:2020mep,Ng:2014pca,Ioka:2014kca,Ibe:2014pja,Kamada:2015era,Shoemaker:2015qul,DiFranzo:2015qea}.
	%A rough estimate shows that a constraint of the order of $\mu^{}_{\nu N} = 10^(-5)~{}~\mu^{}_{\rm B}$ for $m^{}_{N} \lesssim 1~{\rm GeV}$ has already been achieved with the current IceCube observations.
	%We also note that the $\nu^{}_{1}+\gamma^{}_{\rm CMB} \to \nu^{}_{2}$ transition with supernova neutrinos may affect neutrino flavor composition at the Earth. 
	
	\section*{Acknowledgments}
	\noindent GYH is supported by the Alexander von Humboldt Foundation. SJ thanks Ian Shoemaker for useful discussions. 
	
	\begin{widetext}

		\vspace{1cm}
		\appendix
		\section{High-energy neutrino up-scattering cross section}
		\noindent
		Here, we collect the cross section formulae relevant for the conversion of a neutrino to a heavy neutral lepton. 
		Even though the energy of the incoming neutrino (hence also the center-of-mass energy) is extremely high, the conversion process features an enhancement of long-range photon exchange at low momentum transfer, which becomes actually more pronounced with higher incoming neutrino energies. 
		Considering the binding nuclear structure, we distinguish three regimes depending on the magnitude of the momentum transfer $Q^2$~\cite{Magill:2018jla}. At the lowest momentum transfer $0<Q^2<0.5~{\rm GeV^2}$, we will have the coherent elastic scattering off the whole nucleus.
		At the diffractive (quasi-elastic) regime $(0.217~{\rm GeV})^2/A^{2/3}<Q^2<2~{\rm GeV^2}$,
		neutrinos feel and scatter with each individual nucleon in the nucleus (the so-called impulse approximation~\cite{Chew:1952fca}), elastically.
		Those two regimes are experimentally distinguished by their hadronic final state. 
		At large momentum transfer $Q^2>2~{\rm GeV^2}$, the deep inelastic scattering will come into play and we shall calculate the cross section with the parton model.
		It is worth mentioning that the above treatment is also applicable to the trident production of neutrino scatterings~\cite{Magill:2016hgc,Ge:2017poy,Ballett:2018uuc,Zhou:2019vxt}.
		We summarize the formulae for those three regimes below.

		Regardless of which regime we are considering, the cross section can always be formally expressed in terms of electromagnetic form factors (or structure functions). In particular, the squared matrix element factorizes into
		\begin{eqnarray}
		|\mathcal{M}|^2=\frac{\mu^{2}_{ \nu N} e^2}{q^4} L^{}_{\mu \nu}W^{\mu\nu},
		\end{eqnarray}
		where the prefactor accounts for coupling constants and the photon propagator. 
		The leptonic tensor can be straightforwardly obtained  with $L^{}_{\mu\nu} = \sum j^{}_{\mu} j^{\dagger}_{\nu}$, where the lepton current $j^{}_{\mu} = \langle N| \hat{j}^{}_{\mu}|\nu \rangle = 2 \overline{u}^{}_{\nu}(p) P^{}_{\rm R}\sigma^{}_{\mu \alpha} u^{}_{N}(p') q^{\alpha}$ is obtained from Eq.~(\ref{eq:L}).
		The spin-summed leptonic tensor reads
		\begin{eqnarray}
		L^{}_{\mu\nu} = 4 \mathrm{Tr}\left[ \slashed{p} P^{}_{\rm R} \sigma^{}_{\mu\alpha} q^\alpha (\slashed{p}^{\prime}+ m^{}_{N})\sigma^{}_{\nu  \beta}q^\beta \right] .
		\end{eqnarray}
		The general tensor structure of hadronic current can be formally written as~\cite{Drell:1969ca}
		%	\begin{widetext}
		\begin{eqnarray} \label{eq:Wmn}
		W^{}_{\mu\nu} = -\left( g^{}_{\mu\nu} -\frac{q^{}_{\mu}q^{}_{\nu}}{q^2} \right) W^{}_{1}(Q^2,y) + \left(k^{}_{\mu} - \frac{k \cdot q}{q^2}q^{}_{\mu} \right) \left( k^{}_{\nu} - \frac{k \cdot q}{q^2} q^{}_{\nu} \right) \frac{W^{}_{2}(Q^2,y)}{M^2} \;,
		\end{eqnarray}
		%	\end{widetext}
		where the generic structure function $W^{}_{1}$ and $W^{}_{2}$ are both functions of $Q^2$ and $y$.
		%Some remarks are made below for different regimes.
		%\begin{itemize}[noitemsep,topsep=0pt,leftmargin=3.5mm]
		
		\subsection{Elastic regime}
		For elastic scattering (coherent and diffractive regimes), $y$ and $Q^2$ are not independent variables, i.e., $2 M E^{}_{\nu} y/Q^2 = -2k \cdot q/q^2 = 1$. Hence the form factors will be functions of only $Q^2$ or $y$, e.g., $W^{}_{1,2}(Q^2,y) = W^{}_{1,2}(Q^2) $. 
		The hadronic vector current $J^{}_{\mu}$ for a nucleon or nucleus is often parameterized in the general form
		\begin{eqnarray}
		\langle A' | \hat{J}^{}_{\mu} | A \rangle = \overline{u}(k') \left( \gamma^\mu F^{}_{1} + \mathrm{i}\frac{\sigma^{\mu\alpha} q^{}_{\alpha}}{2 M} F^{}_{2} \right) u(k)\;.
		\end{eqnarray}
		In this case, we obtain the form factors in the hadronic tensor $W^{}_{\mu\nu} = \sum j^{\mu}_{\rm H} j^{\dagger \nu}_{\rm H} $ of Eq.~(\ref{eq:Wmn}) to be
		\begin{eqnarray}
		W^{}_{1}(Q^2) &=& 2 Q^2 \left[ F^{}_{1}(Q^2)+F^{}_{2}(Q^2)\right]^2 , \\
		W^{}_{2}(Q^2) &=& 8 M^2 \left[ F^2_{1}(Q^2) + \frac{Q^2}{4M^2}  F^2_{2}(Q^2)  \right] .
		\end{eqnarray}
		The cross section of two-body elastic scattering is trivial and can be derived by integrating over the final-state phase space
		\begin{eqnarray} \label{eq:dsdt}
		\frac{\mathrm{d} \sigma}{\mathrm{d} t} & = &  \frac{2 \alpha  F_1^2 \mu^2_{\rm \mu N}}{t^2 \left(M^2-s\right)^2} \left[ \vphantom{\frac{1}{2}} -2 M^2 m_N^4 - t^2 \left(2 s -2 M^2-m_N^2\right)  -\, t \left(2 s^2 -4 M^2 s -2 m_N^2 s + 2 M^4 + m_N^4 \right) \vphantom{\frac{1}{2}} \right]   \notag\\
		& & -\, \frac{\alpha  F_2^2 \mu^2_{\rm \mu N}}{4 t \left(M^3-M s\right)^2}
		\left[ \vphantom{\frac{1}{2}} 4 M^2 m_N^4 -t^2 \left(4 s- m_N^2\right) -t^3   - \, t \left(4 s^2-8 M^2 s-4  m_N^2 s  + 4 M^4+4 M^2  m_N^2\right) \vphantom{\frac{1}{2}}\right] \notag \\
		& & + \,\frac{2 \alpha  F^{}_1 F^{}_2 \mu^2_{\rm \mu N}}{t \left(M^2- s\right)^2}
		\left( \vphantom{\frac{1}{2}} -2m^4_N + t m^2_{N} + t^2 \right) ,
		\end{eqnarray}
		where the Mandelstam variables $s$ and $t \equiv q^2=-Q^2$ have been adopted. Note that a factor $1/2$ has been considered for averaging over the nuclear spin. The above cross section applies to both coherent and diffractive regimes, and $M$ is the total mass of the nucleus (for coherent one) or nucleon (for diffractive one). 
		The integration limit of momentum transfer is bounded from below by the sterile neutrino mass, namely $Q^2> m^4_N / (4 E^2)$ given $m^2_N \ll 2 M E^{}_{\nu} \ll E^2_{\nu}$.

		For the coherent scattering regime at very low $Q^2$, it is safe to only keep terms with  $F^2_{1}$ given $s \simeq 2M E^{}_{\nu}\gg 2 Q E^{}_{\nu} > m^2_{N}$. Furthermore, the form factor $F^{2}_{1}$ is enhanced by a factor $Z^2$ in the coherent regime, where $Z$ is the atomic number.
		We take the Woods-Saxon form factor such that $F^{}_{1} = Z F^{}_{\rm WS}(Q^2)$, where $F^{}_{\rm WS}(Q^2)$ is parameterized as
		\begin{equation}
		F^{}_{\rm WS} =  \frac{3 \pi  a}{r_0^2 + \pi^2 a^2} \frac{\pi a \coth{(\pi Q a)} \sin{(Q r_0)} - r_0 \cos{(Q r_0)} }{Q r_0 \sinh{(\pi Q a)}}\, .
		\end{equation}
		Here, $a = 0.523~{\rm fm}$ and $r^{}_{0} = 1.126 A^{1/3}~{\rm fm}$ with $A$ being the mass number.
		
		\subsection{Diffractive regime}
		A different parametrization of form factors should be used for the diffractive regime. Following Ref.~\cite{Magill:2018jla} we adopt the following parametrization
		\begin{eqnarray}
		G^{\{\rm p,n \}}_{\rm E} & = & F^{\{\rm p,n \}}_1 - \frac{Q^2}{4 M^2} F^{\{\rm p,n \}}_2 \;, \\
		G^{\{\rm p,n \}}_{\rm M} & = & F^{\{\rm p,n \}}_1 + F^{\{\rm p,n \}}_2 \;, 
		\end{eqnarray}
		where $G^{\{\rm p,n \}}_{\rm E} = \{G^{}_{D},0\}$ and $G^{\{\rm p,n \}}_{\rm M} = \mu^{\{\rm p,n \}} G^{}_{D}$ with $G^{}_{D} = 1/(1+Q^2/0.71~{\rm GeV^2})^2$ and $\mu^{\rm p,n} = \{ 2.793, -1.913\}$.
		%For very small momentum transfer, the Pauli blocking effect in the nucleus will become increasingly important, and we deal with it by treating protons and neutrons in the nucleus as an ideal Fermi gas. This corrects the cross section by multiplying a distribution factor 
		%\begin{equation}
		%f (|\bm{q}|) = \begin{cases} \displaystyle
		%\frac{3}{2} \frac{|\bm{q}|}{2 \, k^{}_{\rm F}} - \frac{1}{2} \left( \frac{|\bm{q}|}{2 \, k^{}_{\rm F}} \right)^3 ,\, &\mathrm{if }\;\; |\bm{q}| < 2\, k^{}_{\rm F}\, ,\\
		%1,\, &\mathrm{if }\;\; |\bm{q}| \geq 2 \, k^{}_{\rm F}\, ,
		%\end{cases}
		%\end{equation}
		%where $k^{}_{\rm F} = 235~{\rm MeV}$.
		
		\subsection{Deep-inelastic scattering regime}
		Now we move to the DIS regime, when $Q^2>2~{\rm GeV}^2$ is satisfied.
		The cross section for $\nu \to N$ conversion is dominated by small momentum transfers.
		Even though $Q^2$ for DIS is large compared to the elastic regimes, the cross section will be enhanced by the parton density which increases with energy scales. Hence, the DIS regime should be included in our calculation. 
		The convention for DIS parameters is as usual:
		\begin{eqnarray}
		\hat{s} & \equiv & 2 M E^{}_{\nu} x + M^2 x^2 \;, \\
		Q^2 & \equiv & 2 M E^{}_{\nu} x y \;,
		\end{eqnarray}
		where $x$ is the Bjorken scaling variable, $y$ is the inelasticity parameter which also measures the energy loss fraction, and $\hat{s}$ is the center-of-mass energy of the incoming neutrino and the parton. 
		
		The hadronic current is simply given by quark partons
		\begin{eqnarray}
		\langle A' | \hat{J}^{}_{{\rm i},\mu} | A \rangle = e^{}_{\rm i}  \, \overline{u}(k') \gamma^\mu  u(x k) \;,
		\end{eqnarray}
		where $e^{}_{\rm i}$ is the charge of the corresponding quark ${\rm q}^{}_{i}=\{\rm u,d,c,s,b \}$.
		It is easy to find  that the structure functions in Eq.~(\ref{eq:Wmn}) are correlated with PDFs by
		\begin{eqnarray}
		W^{}_{1} = \sum_{i}\frac{e^{2}_{i} \cdot {\rm q}^{}_{i}(x)}{M} \;, \, W^{}_{2} =   \sum_{i} \frac{ e^{2}_{i} \cdot 2x{\rm q}^{}_{i}(x)}{y E^{}_{\nu} } \;.
		\end{eqnarray}
		The differential cross section can then be conveniently obtained with
		\begin{eqnarray}
		\frac{\mathrm{d}^2 \sigma}{\mathrm{d}x \mathrm{d}y}  =\frac{2\pi M E y}{E'} \frac{\mathrm{d}^2 \sigma}{\mathrm{d}E' \mathrm{d} \Omega'}\;, \hspace{0.6cm}
		\frac{\mathrm{d}^2 \sigma}{\mathrm{d}E' \mathrm{d} \Omega'}  =  \frac{1}{16\pi^2}\frac{E'}{E} \left(\frac{1}{2} \frac{\mu^{2}_{\rm \mu N} e^2}{q^4} L^{}_{\mu \nu}W^{\mu\nu} \right) .
		\end{eqnarray}
		Another equivalent method is to calculate the parton-level cross section and then convolve with the quark PDFs:
		\begin{eqnarray}
		\frac{\mathrm{d}^2 \sigma}{\mathrm{d}x \mathrm{d}y} & = &2MEx  \frac{\mathrm{d} \sigma(x)}{\mathrm{d} t} \cdot \sum e^2_{i}  {\rm q}^{}_{i}(x)\;, \\
		\frac{\mathrm{d} \sigma(x)}{\mathrm{d} t} & = &  \frac{2 \alpha   \mu^2_{\rm \mu N}}{t^2 \left(x^2 M^2- \hat{s}\right)^2} \left[ \vphantom{\frac{1}{2}} -2 x^2 M^2 m_N^4 - t^2 (2 \hat{s}  - \, 2 x^2 M^2-m_N^2) \right.  \notag\\
		& & \hspace{0.5cm} -\, t (2 \hat{s}^2 -4 x^2M^2 \hat{s}  - \left. 2 m_N^2 \hat{s} + 2 x^4M^4 + m_N^4 ) \vphantom{\frac{1}{2}} \right] ,
		\end{eqnarray}
		where the parton-level cross section $\mathrm{d}\sigma(x)/\mathrm{d}t$ can be directly derived by setting $F^{}_{1} = 1$ and $F^{}_{2}=0$ in Eq.~(\ref{eq:dsdt}) and properly including the scaling parameter.
		The integration should be performed over $x$ and $y$ in the kinematically allowed range 
		%	\begin{widetext}
		\begin{scriptsize}
			\begin{eqnarray}
			\frac{E^{}_{\nu}-\sqrt{E^{2}_{\nu}- m_N^2}}{M} & \leq x \leq  & 1 \;, \\
			\frac{\left(\hat{s} - M^2x^2\right) \left(\hat{s}-\lambda^{1/2}(\hat{s},m^{}_{N},Mx)-M^2 x^2\right)-M^2_N \left(M^2 x^2+\hat{s}\right)}{2 \hat{s} (\hat{s}-M^2 x^2)} & \leq y \leq & \frac{\left(\hat{s} - M^2x^2\right) \left(\hat{s}+\lambda^{1/2}(\hat{s},m^{}_{N},Mx)-M^2 x^2\right)-M^2_N \left(M^2 x^2+\hat{s}\right)}{2 \hat{s} (\hat{s}-M^2 x^2)} \;, \notag
			\end{eqnarray}
		\end{scriptsize}
		%	\end{widetext}
		where $\lambda(x,y,z)\equiv (x-(y-z)^2)(x-(y+z)^2)$.
		The above expression of integration limits is strict and can be approximated to
		\begin{eqnarray}
		\frac{m^2_{N}}{2M E^{}_{\nu}}\leq x \leq  1 \;,~
		\frac{m^4_{N}}{8M E^{3}_{\nu}x}  \leq y \leq  1 \; \notag
		\end{eqnarray}
		providing $m^2_{N} \ll 2ME^{}_{\nu}x\ll E^2_{\nu}$.
		We further confine the integration by requiring $Q^2 = 2MEx y>2~{\rm GeV}$.

		\section{More details on the IceCube fit and tau neutrino telescope configurations}
		\noindent
		Though our main focus in this work is on tau neutrino telescopes, the IceCube result is given for indicative purpose.
		We directly fit to the existing IceCube constraints on neutrino cross sections instead of simulating the events and experimental details ourselves.
		The information we use is the cross section scaling parameter measured in four energy bins, given by Table II of Ref.~\cite{IceCube:2020rnc} (with the frequentist approach): $x^{}_{0} = 0.48^{+0.49}_{-0.37}$ from 60~TeV to 100~TeV, $x^{}_{1} = 1.50^{+1.03}_{-0.60}$ from 100~TeV to 200~TeV, $x^{}_{2} = 0.54^{+0.60}_{-0.35}$ from 200~TeV to 500~TeV and $x^{}_{3} = 2.44^{+5.10}_{-1.47}$ from 500~TeV to 100~PeV.
		The scaling parameter $x^{}_{i}$ describes the ratio of measured cross section to the standard calculation. It is assumed in Ref.~\cite{IceCube:2020rnc} that the CC and NC interactions  vary by the same scaling parameter, which does not necessarily hold for generic new physics modifications.
		In fact, in our scenario with $\nu \to N, N \to \nu \gamma$ only the neutral-current interaction is modified.
		Hence, our fit with IceCube results here is approximative and given just for indication of IceCube constraining power. 
		
		%If we ignore this regeneration effect, it makes basically no difference whether the additional scattering process is CC or NC. Hence, the accuracy of the approximated fit is subject to the impact of regeneration effect, which accounts for .
		%

		We estimate the IceCube constraint as follows.
		For each choice of $m^{}_{N}$ and $\mu^{}_{\nu N}$, we generate the cross sections at the central values of the above energy ranges (in the logarithmic scale), and then fit to the scaling parameter by using the chi-square
		\begin{eqnarray}
		\chi^2_{\rm IC}= \sum^{4}_{i = 1} \frac{\left[ x^{\rm th}_{i} (m^{}_{N},\mu^{}_{\nu N}) -x^{\rm meas}_{i} \right]^2}{(\sigma^{\rm meas}_{x_{i}})^2} \;,
		\end{eqnarray}
		where $x^{\rm meas}_{i}$ is the central value of the scaling parameter, and $\sigma^{\rm meas}_{x_{i}}$ is the $1\sigma$ uncertainty. The asymmetric error $\sigma^{\rm meas}_{x_{i}}$ is chosen depending on the sign of $x^{\rm th}_{i}-x^{\rm meas}_{i}$.

		Let us comment on the limitation of the simplified treatment instead of a full simulation of IceCube events.
		The main effect of a modified neutrino cross section is to attenuate the neutrino flux before reaching the detector. The reconstruction of the cross section strongly relies on the angular distribution of neutrino events. 
		The attenuation effect can be approximately parameterized as $\Phi^{\rm out}_{\nu} = \Phi^{\rm in}_{\nu} \mathrm{e}^{-\sigma \cdot n \cdot L}$, where $\sigma$ accounts for both CC and NC interactions, and $n$ is the target density.
		However, the production of secondary neutrinos from NC and tau decays (for $\nu^{}_{\tau}$ CC only) will severely compensate the neutrino attenuation. For the NC process, on average $80\%$ of the incoming neutrino energy will flow into the secondary neutrino, which effectively reduces the neutrino scattering cross section.
		The tau regeneration will result in a secondary neutrino with approximately $30\%$ of the initial neutrino energy on average.
		As a consequence, the outcome of a neutrino scattering not only depends on whether it is CC or NC but also on the neutrino flavor, and hence a general modification to the cross section cannot be fully represented by the scaling parameters $x^{}_{i}$.
		Nevertheless, we assume that three generations of neutrinos share the same magnetic dipole moment for the IceCube estimate, and calculate the theoretical expectations by comparing the $\nu \to N$ cross section to the total cross section of SM CC and NC processes. According to the considerations above, our estimate can only serve as an indication of the IceCube capability.
		However, since the current IceCube limits on $x^{}_{i}$ are still very loose, e.g., the $1\sigma$ interval of $x^{}_{0}$ ranges from $0.1$ to $0.9$ (by almost one order of magnitude), we do not expect significant changes with a more dedicated IceCube data analysis.
		
		We give more details on how we obtain the sensitivity of three representative tau neutrino telescopes, GRAND, POEMMA and Trinity. We follow closely the technical details described in Ref.~\cite{Huang:2021mki}. In the following, we make a brief summary.
		After the outgoing tau flux for given matter profiles is obtained with the help of Eqs.~(\ref{eq:nupropa1}) and ~(\ref{eq:nupropa2}), we integrate it with the actual experimental configurations. 
		For GRAND, we take 20 copies of 10k radio antenna arrays and assume that these antennas of each copy form a projection screen with an area of $150~{\rm km} \times 66~{\rm km}$. The screen is assumed to be deployed on a mountain slope inclined by $3^{\circ}$ from the horizontal, and overwatch another mountain with a height of $2.5~{\rm km}$ and a width of $40~{\rm km}$. The distance between two mountains over the valley is set to $10~{\rm km}$.
		For POEMMA, we assume two satellites orbiting with an height of $525~{\rm km}$. Each of the satellite carries an optical camera with an area $5.71~{\rm m^2}$ on-axis.
		For Trinity, three stations of  are assumed, and each of them has an imaging system with $360^{\circ}$ field of view in azimuth and $5^{\circ}$ in zenith. These stations are placed on the top of a mountain with a height of $2~{\rm km}$.
	\end{widetext}
	
	\bibliographystyle{utcaps_mod}
	\bibliography{references}

\providecommand{\href}[2]{#2}\begingroup\raggedright\begin{thebibliography}{100}

\bibitem{Athar:2021xsd}
M.~S. Athar {\em et al.}, ``{\em {Status and perspectives of neutrino
  physics}},'' \href{http://dx.doi.org/10.1016/j.ppnp.2022.103947}{Prog. Part.
  Nucl. Phys. {\normalfont \bfseries 124} (2022)  103947},
  \href{http://arxiv.org/abs/2111.07586}{{\normalfont \ttfamily
  arXiv:2111.07586}}.

\bibitem{ParticleDataGroup:2020ssz}
{\normalfont \bfseries Particle Data Group}, P.~A. Zyla {\em et al.}, ``{\em
  {Review of Particle Physics}},''
  \href{http://dx.doi.org/10.1093/ptep/ptaa104}{PTEP {\normalfont \bfseries
  2020} (2020) no.~8, 083C01}.

\bibitem{Cowan:1954pq}
C.~L. Cowan, F.~Reines, and F.~B. Harrison, ``{\em {Upper limit on the neutrino
  magnetic moment}},'' \href{http://dx.doi.org/10.1103/PhysRev.96.1294}{Phys.
  Rev. {\normalfont \bfseries 96} (1954)  1294}.

\bibitem{Davis:1988gd}
R.~Davis, K.~Lande, B.~T. Cleveland, J.~K. Rowley, and J.~Ullman, ``{\em
  {Report on the chlorine solar neutrino experiment}},'' Conf. Proc. C
  {\normalfont \bfseries 880605} (1988)  518--525.

\bibitem{Davis:1990fb}
R.~Davis, K.~Lande, C.~K. Lee, B.~T. Cleveland, and J.~Ullman, ``{\em {Results
  of the Homestake chlorine solar neutrino experiment}},'' in {\em {21st
  International Cosmic Ray Conference}}, pp.~155--158.
\newblock 1990.

\bibitem{Vidyakin:1992nf}
G.~S. Vidyakin, V.~N. Vyrodov, I.~I. Gurevich, Y.~V. Kozlov, V.~P. Martemyanov,
  S.~V. Sukhotin, V.~G. Tarasenkov, E.~V. Turbin, and S.~K. Khakhimov, ``{\em
  {Limitations on the magnetic moment and charge radius of the
  electron-anti-neutrino}},'' JETP Lett. {\normalfont \bfseries 55} (1992)
  206--210.

\bibitem{Derbin:1993wy}
A.~I. Derbin, A.~V. Chernyi, L.~A. Popeko, V.~N. Muratova, G.~A. Shishkina, and
  S.~I. Bakhlanov, ``{\em {Experiment on anti-neutrino scattering by electrons
  at a reactor of the Rovno nuclear power plant}},'' JETP Lett. {\normalfont
  \bfseries 57} (1993)  768--772.

\bibitem{LSND:2001akn}
{\normalfont \bfseries LSND}, L.~B. Auerbach {\em et al.}, ``{\em {Measurement
  of electron - neutrino - electron elastic scattering}},''
  \href{http://dx.doi.org/10.1103/PhysRevD.63.112001}{Phys. Rev. D {\normalfont
  \bfseries 63} (2001)  112001},
  \href{http://arxiv.org/abs/hep-ex/0101039}{{\normalfont \ttfamily
  arXiv:hep-ex/0101039}}.

\bibitem{Daraktchieva:2005kn}
{\normalfont \bfseries MUNU}, Z.~Daraktchieva {\em et al.}, ``{\em {Final
  results on the neutrino magnetic moment from the MUNU experiment}},''
  \href{http://dx.doi.org/10.1016/j.physletb.2005.04.030}{Phys. Lett. B
  {\normalfont \bfseries 615} (2005)  153--159},
  \href{http://arxiv.org/abs/hep-ex/0502037}{{\normalfont \ttfamily
  arXiv:hep-ex/0502037}}.

\bibitem{Deniz:2009mu}
{\normalfont \bfseries TEXONO}, M.~Deniz {\em et al.}, ``{\em {Measurement of
  Nu(e)-bar -Electron Scattering Cross-Section with a CsI(Tl) Scintillating
  Crystal Array at the Kuo-Sheng Nuclear Power Reactor}},''
  \href{http://dx.doi.org/10.1103/PhysRevD.81.072001}{Phys. Rev. D {\normalfont
  \bfseries 81} (2010)  072001},
  \href{http://arxiv.org/abs/0911.1597}{{\normalfont \ttfamily
  arXiv:0911.1597}}.

\bibitem{Beda:2012zz}
A.~G. Beda, V.~B. Brudanin, V.~G. Egorov, D.~V. Medvedev, V.~S. Pogosov, M.~V.
  Shirchenko, and A.~S. Starostin, ``{\em {The results of search for the
  neutrino magnetic moment in GEMMA experiment}},''
  \href{http://dx.doi.org/10.1155/2012/350150}{Adv. High Energy Phys.
  {\normalfont \bfseries 2012} (2012)  350150}.

\bibitem{Allen:1992qe}
R.~C. Allen {\em et al.}, ``{\em {Study of electron-neutrino electron elastic
  scattering at LAMPF}},''
  \href{http://dx.doi.org/10.1103/PhysRevD.47.11}{Phys. Rev. D {\normalfont
  \bfseries 47} (1993)  11--28}.

\bibitem{Borexino:2017fbd}
{\normalfont \bfseries Borexino}, M.~Agostini {\em et al.}, ``{\em {Limiting
  neutrino magnetic moments with Borexino Phase-II solar neutrino data}},''
  \href{http://dx.doi.org/10.1103/PhysRevD.96.091103}{Phys. Rev. D {\normalfont
  \bfseries 96} (2017) no.~9, 091103},
  \href{http://arxiv.org/abs/1707.09355}{{\normalfont \ttfamily
  arXiv:1707.09355}}.

\bibitem{Bonet:2022imz}
{\normalfont \bfseries CONUS}, H.~Bonet {\em et al.}, ``{\em {First limits on
  neutrino electromagnetic properties from the CONUS experiment}},''
  \href{http://arxiv.org/abs/2201.12257}{{\normalfont \ttfamily
  arXiv:2201.12257}}.

\bibitem{Aprile:2020tmw}
{\normalfont \bfseries XENON}, E.~Aprile {\em et al.}, ``{\em {Excess
  electronic recoil events in XENON1T}},''
  \href{http://dx.doi.org/10.1103/PhysRevD.102.072004}{Phys. Rev. D
  {\normalfont \bfseries 102} (2020) no.~7, 072004},
  \href{http://arxiv.org/abs/2006.09721}{{\normalfont \ttfamily
  arXiv:2006.09721}}.

\bibitem{ANITA:2016vrp}
{\normalfont \bfseries ANITA}, P.~W. Gorham {\em et al.}, ``{\em
  {Characteristics of Four Upward-pointing Cosmic-ray-like Events Observed with
  ANITA}},'' \href{http://dx.doi.org/10.1103/PhysRevLett.117.071101}{Phys. Rev.
  Lett. {\normalfont \bfseries 117} (2016) no.~7, 071101},
  \href{http://arxiv.org/abs/1603.05218}{{\normalfont \ttfamily
  arXiv:1603.05218}}.

\bibitem{ANITA:2018sgj}
{\normalfont \bfseries ANITA}, P.~W. Gorham {\em et al.}, ``{\em {Observation
  of an Unusual Upward-going Cosmic-ray-like Event in the Third Flight of
  ANITA}},'' \href{http://dx.doi.org/10.1103/PhysRevLett.121.161102}{Phys. Rev.
  Lett. {\normalfont \bfseries 121} (2018) no.~16, 161102},
  \href{http://arxiv.org/abs/1803.05088}{{\normalfont \ttfamily
  arXiv:1803.05088}}.

\bibitem{MiniBooNE:2018esg}
{\normalfont \bfseries MiniBooNE}, A.~A. Aguilar-Arevalo {\em et al.}, ``{\em
  {Significant Excess of ElectronLike Events in the MiniBooNE Short-Baseline
  Neutrino Experiment}},''
  \href{http://dx.doi.org/10.1103/PhysRevLett.121.221801}{Phys. Rev. Lett.
  {\normalfont \bfseries 121} (2018) no.~22, 221801},
  \href{http://arxiv.org/abs/1805.12028}{{\normalfont \ttfamily
  arXiv:1805.12028}}.

\bibitem{Muong-2:2021ojo}
{\normalfont \bfseries Muon g-2}, B.~Abi {\em et al.}, ``{\em {Measurement of
  the Positive Muon Anomalous Magnetic Moment to 0.46 ppm}},''
  \href{http://dx.doi.org/10.1103/PhysRevLett.126.141801}{Phys. Rev. Lett.
  {\normalfont \bfseries 126} (2021) no.~14, 141801},
  \href{http://arxiv.org/abs/2104.03281}{{\normalfont \ttfamily
  arXiv:2104.03281}}.

\bibitem{Babu:2021jnu}
K.~S. Babu, S.~Jana, M.~Lindner, and V.~P. K, ``{\em {Muon g \ensuremath{-} 2
  anomaly and neutrino magnetic moments}},''
  \href{http://dx.doi.org/10.1007/JHEP10(2021)240}{JHEP {\normalfont \bfseries
  10} (2021)  240}, \href{http://arxiv.org/abs/2104.03291}{{\normalfont
  \ttfamily arXiv:2104.03291}}.

\bibitem{Ismail:2021dyp}
A.~Ismail, S.~Jana, and R.~M. Abraham, ``{\em {Neutrino up-scattering via the
  dipole portal at forward LHC detectors}},''
  \href{http://dx.doi.org/10.1103/PhysRevD.105.055008}{Phys. Rev. D
  {\normalfont \bfseries 105} (2022) no.~5, 055008},
  \href{http://arxiv.org/abs/2109.05032}{{\normalfont \ttfamily
  arXiv:2109.05032}}.

\bibitem{Magill:2018jla}
G.~Magill, R.~Plestid, M.~Pospelov, and Y.-D. Tsai, ``{\em {Dipole Portal to
  Heavy Neutral Leptons}},''
  \href{http://dx.doi.org/10.1103/PhysRevD.98.115015}{Phys. Rev. D {\normalfont
  \bfseries 98} (2018) no.~11, 115015},
  \href{http://arxiv.org/abs/1803.03262}{{\normalfont \ttfamily
  arXiv:1803.03262}}.

\bibitem{Gninenko:2009ks}
S.~N. Gninenko, ``{\em {The MiniBooNE anomaly and heavy neutrino decay}},''
  \href{http://dx.doi.org/10.1103/PhysRevLett.103.241802}{Phys. Rev. Lett.
  {\normalfont \bfseries 103} (2009)  241802},
  \href{http://arxiv.org/abs/0902.3802}{{\normalfont \ttfamily
  arXiv:0902.3802}}.

\bibitem{Schwetz:2020xra}
T.~Schwetz, A.~Zhou, and J.-Y. Zhu, ``{\em {Constraining active-sterile
  neutrino transition magnetic moments at DUNE near and far detectors}},''
  \href{http://dx.doi.org/10.1007/JHEP07(2021)200}{JHEP {\normalfont \bfseries
  21} (2020)  200}, \href{http://arxiv.org/abs/2105.09699}{{\normalfont
  \ttfamily arXiv:2105.09699}}.

\bibitem{Brdar:2020quo}
V.~Brdar, A.~Greljo, J.~Kopp, and T.~Opferkuch, ``{\em {The Neutrino Magnetic
  Moment Portal: Cosmology, Astrophysics, and Direct Detection}},''
  \href{http://dx.doi.org/10.1088/1475-7516/2021/01/039}{JCAP {\normalfont
  \bfseries 01} (2021)  039},
  \href{http://arxiv.org/abs/2007.15563}{{\normalfont \ttfamily
  arXiv:2007.15563}}.

\bibitem{Shoemaker:2020kji}
I.~M. Shoemaker, Y.-D. Tsai, and J.~Wyenberg, ``{\em {Active-to-sterile
  neutrino dipole portal and the XENON1T excess}},''
  \href{http://dx.doi.org/10.1103/PhysRevD.104.115026}{Phys. Rev. D
  {\normalfont \bfseries 104} (2021) no.~11, 115026},
  \href{http://arxiv.org/abs/2007.05513}{{\normalfont \ttfamily
  arXiv:2007.05513}}.

\bibitem{Bolton:2021pey}
P.~D. Bolton, F.~F. Deppisch, K.~Fridell, J.~Harz, C.~Hati, and S.~Kulkarni,
  ``{\em {Probing Active-Sterile Neutrino Transition Magnetic Moments with
  Photon Emission from CE$\nu$NS}},''
  \href{http://arxiv.org/abs/2110.02233}{{\normalfont \ttfamily
  arXiv:2110.02233}}.

\bibitem{Miranda:2021kre}
O.~G. Miranda, D.~K. Papoulias, O.~Sanders, M.~T\'ortola, and J.~W.~F. Valle,
  ``{\em {Low-energy probes of sterile neutrino transition magnetic
  moments}},'' \href{http://dx.doi.org/10.1007/JHEP12(2021)191}{JHEP
  {\normalfont \bfseries 12} (2021)  191},
  \href{http://arxiv.org/abs/2109.09545}{{\normalfont \ttfamily
  arXiv:2109.09545}}.

\bibitem{Vergani:2021tgc}
S.~Vergani, N.~W. Kamp, A.~Diaz, C.~A. Arg\"uelles, J.~M. Conrad, M.~H.
  Shaevitz, and M.~A. Uchida, ``{\em {Explaining the MiniBooNE excess through a
  mixed model of neutrino oscillation and decay}},''
  \href{http://dx.doi.org/10.1103/PhysRevD.104.095005}{Phys. Rev. D
  {\normalfont \bfseries 104} (2021) no.~9, 095005},
  \href{http://arxiv.org/abs/2105.06470}{{\normalfont \ttfamily
  arXiv:2105.06470}}.

\bibitem{Zhang:2022spf}
Y.~Zhang, M.~Song, R.~Ding, and L.~Chen, ``{\em {Neutrino dipole portal at
  electron colliders}},'' \href{http://arxiv.org/abs/2204.07802}{{\normalfont
  \ttfamily arXiv:2204.07802}}.

\bibitem{Raffelt:1999tx}
G.~G. Raffelt, ``{\em {Particle physics from stars}},''
  \href{http://dx.doi.org/10.1146/annurev.nucl.49.1.163}{Ann. Rev. Nucl. Part.
  Sci. {\normalfont \bfseries 49} (1999)  163--216},
  \href{http://arxiv.org/abs/hep-ph/9903472}{{\normalfont \ttfamily
  arXiv:hep-ph/9903472}}.

\bibitem{Capozzi:2020cbu}
F.~Capozzi and G.~Raffelt, ``{\em {Axion and neutrino bounds improved with new
  calibrations of the tip of the red-giant branch using geometric distance
  determinations}},''
  \href{http://dx.doi.org/10.1103/PhysRevD.102.083007}{Phys. Rev. D
  {\normalfont \bfseries 102} (2020) no.~8, 083007},
  \href{http://arxiv.org/abs/2007.03694}{{\normalfont \ttfamily
  arXiv:2007.03694}}.

\bibitem{Jana:2022tsa}
S.~Jana, Y.~P. Porto-Silva, and M.~Sen, ``{\em {Exploiting a future galactic
  supernova to probe neutrino magnetic moments}},''
  \href{http://arxiv.org/abs/2203.01950}{{\normalfont \ttfamily
  arXiv:2203.01950}}.

\bibitem{Fujikawa:1980yx}
K.~Fujikawa and R.~Shrock, ``{\em {The Magnetic Moment of a Massive Neutrino
  and Neutrino Spin Rotation}},''
  \href{http://dx.doi.org/10.1103/PhysRevLett.45.963}{Phys. Rev. Lett.
  {\normalfont \bfseries 45} (1980)  963}.

\bibitem{Pal:1981rm}
P.~B. Pal and L.~Wolfenstein, ``{\em {Radiative Decays of Massive
  Neutrinos}},'' \href{http://dx.doi.org/10.1103/PhysRevD.25.766}{Phys. Rev. D
  {\normalfont \bfseries 25} (1982)  766}.

\bibitem{Giunti:2014ixa}
C.~Giunti and A.~Studenikin, ``{\em {Neutrino electromagnetic interactions: a
  window to new physics}},''
  \href{http://dx.doi.org/10.1103/RevModPhys.87.531}{Rev. Mod. Phys.
  {\normalfont \bfseries 87} (2015)  531},
  \href{http://arxiv.org/abs/1403.6344}{{\normalfont \ttfamily
  arXiv:1403.6344}}.

\bibitem{Babu:2020ivd}
K.~S. Babu, S.~Jana, and M.~Lindner, ``{\em {Large Neutrino Magnetic Moments in
  the Light of Recent Experiments}},''
  \href{http://dx.doi.org/10.1007/JHEP10(2020)040}{JHEP {\normalfont \bfseries
  10} (2020)  040}, \href{http://arxiv.org/abs/2007.04291}{{\normalfont
  \ttfamily arXiv:2007.04291}}.

\bibitem{Babu:2019mfe}
K.~S. Babu, P.~S.~B. Dev, S.~Jana, and A.~Thapa, ``{\em {Non-Standard
  Interactions in Radiative Neutrino Mass Models}},''
  \href{http://dx.doi.org/10.1007/JHEP03(2020)006}{JHEP {\normalfont \bfseries
  03} (2020)  006}, \href{http://arxiv.org/abs/1907.09498}{{\normalfont
  \ttfamily arXiv:1907.09498}}.

\bibitem{Voloshin:1987qy}
M.~B. Voloshin, ``{\em {On Compatibility of Small Mass with Large Magnetic
  Moment of Neutrino}},'' Sov. J. Nucl. Phys. {\normalfont \bfseries 48} (1988)
   512.

\bibitem{Babu:1989wn}
K.~S. Babu and R.~N. Mohapatra, ``{\em {Model for Large Transition Magnetic
  Moment of the $\nu_e$}},''
  \href{http://dx.doi.org/10.1103/PhysRevLett.63.228}{Phys. Rev. Lett.
  {\normalfont \bfseries 63} (1989)  228}.

\bibitem{Barr:1990um}
S.~M. Barr, E.~M. Freire, and A.~Zee, ``{\em {A Mechanism for large neutrino
  magnetic moments}},''
  \href{http://dx.doi.org/10.1103/PhysRevLett.65.2626}{Phys. Rev. Lett.
  {\normalfont \bfseries 65} (1990)  2626--2629}.

\bibitem{Primakoff:1951iae}
H.~Primakoff, ``{\em {Photoproduction of neutral mesons in nuclear electric
  fields and the mean life of the neutral meson}},''
  \href{http://dx.doi.org/10.1103/PhysRev.81.899}{Phys. Rev. {\normalfont
  \bfseries 81} (1951)  899}.

\bibitem{IceCube:2013cdw}
{\normalfont \bfseries IceCube}, M.~G. Aartsen {\em et al.}, ``{\em {First
  observation of PeV-energy neutrinos with IceCube}},''
  \href{http://dx.doi.org/10.1103/PhysRevLett.111.021103}{Phys. Rev. Lett.
  {\normalfont \bfseries 111} (2013)  021103},
  \href{http://arxiv.org/abs/1304.5356}{{\normalfont \ttfamily
  arXiv:1304.5356}}.

\bibitem{Ogawa:2021dK}
S.~Ogawa and M.~Sasaki, ``{\em {Galactic Bulge VHE tau-neutrino and gamma-ray
  Monitor with Ashra-1 and NTA detectors}},''
  \href{http://dx.doi.org/10.22323/1.395.0970}{PoS {\normalfont \bfseries
  ICRC2021} (2021)  970}.

\bibitem{Wissel:2020fav}
S.~Wissel {\em et al.}, ``{\em {Concept Study for the Beamforming Elevated
  Array for Cosmic Neutrinos (BEACON)}},''
  \href{http://dx.doi.org/10.22323/1.358.1033}{PoS {\normalfont \bfseries
  ICRC2019} (2020)  1033}.

\bibitem{Wissel:2020sec}
S.~Wissel {\em et al.}, ``{\em {Prospects for high-elevation radio detection of
  \ensuremath{>}100 PeV tau neutrinos}},''
  \href{http://dx.doi.org/10.1088/1475-7516/2020/11/065}{JCAP {\normalfont
  \bfseries 11} (2020)  065},
  \href{http://arxiv.org/abs/2004.12718}{{\normalfont \ttfamily
  arXiv:2004.12718}}.

\bibitem{GRAND:2018iaj}
{\normalfont \bfseries GRAND}, J.~\'Alvarez-Mu\~niz {\em et al.}, ``{\em {The
  Giant Radio Array for Neutrino Detection (GRAND): Science and Design}},''
  \href{http://dx.doi.org/10.1007/s11433-018-9385-7}{Sci. China Phys. Mech.
  Astron. {\normalfont \bfseries 63} (2020) no.~1, 219501},
  \href{http://arxiv.org/abs/1810.09994}{{\normalfont \ttfamily
  arXiv:1810.09994}}.

\bibitem{Kotera:2021ca}
K.~Kotera, ``{\em {The Giant Radio Array for Neutrino Detection (GRAND)
  project}},'' \href{http://dx.doi.org/10.22323/1.395.1181}{PoS {\normalfont
  \bfseries ICRC2021} (2021)  1181}.

\bibitem{IceCube-Gen2:2020qha}
{\normalfont \bfseries IceCube-Gen2}, M.~G. Aartsen {\em et al.}, ``{\em
  {IceCube-Gen2: the window to the extreme Universe}},''
  \href{http://dx.doi.org/10.1088/1361-6471/abbd48}{J. Phys. G {\normalfont
  \bfseries 48} (2021) no.~6, 060501},
  \href{http://arxiv.org/abs/2008.04323}{{\normalfont \ttfamily
  arXiv:2008.04323}}.

\bibitem{Hallmann:2021kqk}
{\normalfont \bfseries IceCube-Gen2}, R.~Abbasi {\em et al.}, ``{\em
  {Sensitivity studies for the IceCube-Gen2 radio array}},''
  \href{http://dx.doi.org/10.22323/1.395.1183}{PoS {\normalfont \bfseries
  ICRC2021} (2021)  1183}, \href{http://arxiv.org/abs/2107.08910}{{\normalfont
  \ttfamily arXiv:2107.08910}}.

\bibitem{POEMMA:2020ykm}
{\normalfont \bfseries POEMMA}, A.~V. Olinto {\em et al.}, ``{\em {The POEMMA
  (Probe of Extreme Multi-Messenger Astrophysics) observatory}},''
  \href{http://dx.doi.org/10.1088/1475-7516/2021/06/007}{JCAP {\normalfont
  \bfseries 06} (2021)  007},
  \href{http://arxiv.org/abs/2012.07945}{{\normalfont \ttfamily
  arXiv:2012.07945}}.

\bibitem{Romero-Wolf:2020pzh}
A.~Romero-Wolf {\em et al.}, ``{\em {An Andean Deep-Valley Detector for
  High-Energy Tau Neutrinos}},'' in {\em {Latin American Strategy Forum for
  Research Infrastructure}}.
\newblock 2, 2020.
\newblock \href{http://arxiv.org/abs/2002.06475}{{\normalfont \ttfamily
  arXiv:2002.06475}}.

\bibitem{Otte:2018uxj}
A.~N. Otte, ``{\em {Studies of an air-shower imaging system for the detection
  of ultrahigh-energy neutrinos}},''
  \href{http://dx.doi.org/10.1103/PhysRevD.99.083012}{Phys. Rev. D {\normalfont
  \bfseries 99} (2019) no.~8, 083012},
  \href{http://arxiv.org/abs/1811.09287}{{\normalfont \ttfamily
  arXiv:1811.09287}}.

\bibitem{Otte:2019aaf}
A.~N. Otte, A.~M. Brown, M.~Doro, A.~Falcone, J.~Holder, E.~Judd, P.~Kaaret,
  M.~Mariotti, K.~Murase, and I.~Taboada, ``{\em {Trinity: An Air-Shower
  Imaging Instrument to detect Ultrahigh Energy Neutrinos}},''
  \href{http://arxiv.org/abs/1907.08727}{{\normalfont \ttfamily
  arXiv:1907.08727}}.

\bibitem{Wang:2021/M}
A.~Wang, C.~Lin, N.~Otte, M.~Doro, E.~Gazda, I.~Taboada, A.~Brown, and
  M.~Bagheri, ``{\em {Trinity’s Sensitivity to Isotropic and Point-Source
  Neutrinos}},'' \href{http://dx.doi.org/10.22323/1.395.1234}{PoS {\normalfont
  \bfseries ICRC2021} (2021)  1234}.

\bibitem{Brown:2021tf}
A.~Brown, ``{\em {Trinity: an imaging air Cherenkov telescope to search for
  Ultra-High-Energy neutrinos.}},''
  \href{http://dx.doi.org/10.22323/1.395.1179}{PoS {\normalfont \bfseries
  ICRC2021} (2021)  1179}.

\bibitem{Huang:2021mki}
G.-y. Huang, S.~Jana, M.~Lindner, and W.~Rodejohann, ``{\em {Probing new
  physics at future tau neutrino telescopes}},''
  \href{http://dx.doi.org/10.1088/1475-7516/2022/02/038}{JCAP {\normalfont
  \bfseries 02} (2022) no.~02, 038},
  \href{http://arxiv.org/abs/2112.09476}{{\normalfont \ttfamily
  arXiv:2112.09476}}.

\bibitem{Abraham:2022jse}
R.~M. Abraham {\em et al.}, ``{\em {Tau Neutrinos in the Next Decade: from GeV
  to EeV}},'' \href{http://arxiv.org/abs/2203.05591}{{\normalfont \ttfamily
  arXiv:2203.05591}}.

\bibitem{Ackermann:2022rqc}
M.~Ackermann {\em et al.}, ``{\em {High-Energy and Ultra-High-Energy
  Neutrinos}},'' \href{http://arxiv.org/abs/2203.08096}{{\normalfont \ttfamily
  arXiv:2203.08096}}.

\bibitem{Arguelles:2022xxa}
C.~A. Arg\"uelles {\em et al.}, ``{\em {Snowmass White Paper: Beyond the
  Standard Model effects on Neutrino Flavor}},'' in {\em {2022 Snowmass Summer
  Study}}.
\newblock 3, 2022.
\newblock \href{http://arxiv.org/abs/2203.10811}{{\normalfont \ttfamily
  arXiv:2203.10811}}.

\bibitem{Berezinsky:1969qj}
V.~S. Berezinsky and G.~T. Zatsepin, ``{\em {Cosmic rays at ultrahigh-energies
  (neutrino?)}},'' \href{http://dx.doi.org/10.1016/0370-2693(69)90341-4}{Phys.
  Lett. B {\normalfont \bfseries 28} (1969)  423--424}.

\bibitem{Greisen:1966jv}
K.~Greisen, ``{\em {End to the cosmic ray spectrum?}},''
  \href{http://dx.doi.org/10.1103/PhysRevLett.16.748}{Phys. Rev. Lett.
  {\normalfont \bfseries 16} (1966)  748--750}.

\bibitem{Zatsepin:1966jv}
G.~T. Zatsepin and V.~A. Kuzmin, ``{\em {Upper limit of the spectrum of cosmic
  rays}},'' JETP Lett. {\normalfont \bfseries 4} (1966)  78--80.

\bibitem{Jezo:2014kla}
T.~Je\v{z}o, M.~Klasen, F.~Lyonnet, F.~Montanet, I.~Schienbein, and M.~Tartare,
  ``{\em {Can new heavy gauge bosons be observed in ultra-high energy cosmic
  neutrino events?}},''
  \href{http://dx.doi.org/10.1103/PhysRevD.89.077702}{Phys. Rev. D {\normalfont
  \bfseries 89} (2014) no.~7, 077702},
  \href{http://arxiv.org/abs/1401.6012}{{\normalfont \ttfamily
  arXiv:1401.6012}}.

\bibitem{Jho:2018dvt}
Y.~Jho and S.~C. Park, ``{\em {Probing new physics with high-multiplicity
  events: Ultrahigh-energy neutrinos at air-shower detector arrays}},''
  \href{http://dx.doi.org/10.1103/PhysRevD.104.015018}{Phys. Rev. D
  {\normalfont \bfseries 104} (2021) no.~1, 015018},
  \href{http://arxiv.org/abs/1806.03063}{{\normalfont \ttfamily
  arXiv:1806.03063}}.

\bibitem{Huang:2019hgs}
G.-y. Huang and Q.~Liu, ``{\em {Hunting the Glashow Resonance with PeV Neutrino
  Telescopes}},'' \href{http://dx.doi.org/10.1088/1475-7516/2020/03/005}{JCAP
  {\normalfont \bfseries 03} (2020)  005},
  \href{http://arxiv.org/abs/1912.02976}{{\normalfont \ttfamily
  arXiv:1912.02976}}.

\bibitem{Denton:2020jft}
P.~B. Denton and Y.~Kini, ``{\em {Ultra-High-Energy Tau Neutrino Cross Sections
  with GRAND and POEMMA}},''
  \href{http://dx.doi.org/10.1103/PhysRevD.102.123019}{Phys. Rev. D
  {\normalfont \bfseries 102} (2020)  123019},
  \href{http://arxiv.org/abs/2007.10334}{{\normalfont \ttfamily
  arXiv:2007.10334}}.

\bibitem{Valera:2021dix}
V.~B. Valera and M.~Bustamante, ``{\em {Reaching the EeV frontier in
  neutrino-nucleon cross sections in upcoming neutrino telescopes}},''
  \href{http://dx.doi.org/10.22323/1.395.1200}{PoS {\normalfont \bfseries
  ICRC2021} (2021)  1200}.

\bibitem{Soto:2021vdc}
A.~G. Soto, P.~Zhelnin, I.~Safa, and C.~A. Arg\"uelles, ``{\em {Tau Appearance
  from High-Energy Neutrino Interactions}},''
  \href{http://arxiv.org/abs/2112.06937}{{\normalfont \ttfamily
  arXiv:2112.06937}}.

\bibitem{Valera:2022ylt}
V.~B. Valera, M.~Bustamante, and C.~Glaser, ``{\em {The ultra-high-energy
  neutrino-nucleon cross section: measurement forecasts for an era of cosmic
  EeV-neutrino discovery}},''
  \href{http://arxiv.org/abs/2204.04237}{{\normalfont \ttfamily
  arXiv:2204.04237}}.

\bibitem{Xing:2011zm}
Z.-z. Xing and S.~Zhou, ``{\em {The Glashow resonance as a discriminator of UHE
  cosmic neutrinos originating from p-gamma and p-p collisions}},''
  \href{http://dx.doi.org/10.1103/PhysRevD.84.033006}{Phys. Rev. D {\normalfont
  \bfseries 84} (2011)  033006},
  \href{http://arxiv.org/abs/1105.4114}{{\normalfont \ttfamily
  arXiv:1105.4114}}.

\bibitem{Babu:2019vff}
K.~S. Babu, P.~S. Dev, S.~Jana, and Y.~Sui, ``{\em {Zee-Burst: A New Probe of
  Neutrino Nonstandard Interactions at IceCube}},''
  \href{http://dx.doi.org/10.1103/PhysRevLett.124.041805}{Phys. Rev. Lett.
  {\normalfont \bfseries 124} (2020) no.~4, 041805},
  \href{http://arxiv.org/abs/1908.02779}{{\normalfont \ttfamily
  arXiv:1908.02779}}.

\bibitem{Bustamante:2020niz}
M.~Bustamante, ``{\em {New limits on neutrino decay from the Glashow resonance
  of high-energy cosmic neutrinos}},''
  \href{http://arxiv.org/abs/2004.06844}{{\normalfont \ttfamily
  arXiv:2004.06844}}.

\bibitem{Zhou:2020oym}
S.~Zhou, ``{\em {Cosmic Flavor Hexagon for Ultrahigh-energy Neutrinos and
  Antineutrinos at Neutrino Telescopes}},''
  \href{http://arxiv.org/abs/2006.06181}{{\normalfont \ttfamily
  arXiv:2006.06181}}.

\bibitem{Song:2020nfh}
N.~Song, S.~W. Li, C.~A. Arg\"uelles, M.~Bustamante, and A.~C. Vincent, ``{\em
  {The Future of High-Energy Astrophysical Neutrino Flavor Measurements}},''
  \href{http://dx.doi.org/10.1088/1475-7516/2021/04/054}{JCAP {\normalfont
  \bfseries 04} (2021)  054},
  \href{http://arxiv.org/abs/2012.12893}{{\normalfont \ttfamily
  arXiv:2012.12893}}.

\bibitem{Dey:2020fbx}
U.~K. Dey, N.~Nath, and S.~Sadhukhan, ``{\em {Charged Higgs effects in IceCube:
  PeV events and NSIs}},''
  \href{http://dx.doi.org/10.1007/JHEP09(2021)113}{JHEP {\normalfont \bfseries
  09} (2021)  113}, \href{http://arxiv.org/abs/2010.05797}{{\normalfont
  \ttfamily arXiv:2010.05797}}.

\bibitem{Dev:2016uxj}
P.~S.~B. Dev, D.~K. Ghosh, and W.~Rodejohann, ``{\em {R-parity Violating
  Supersymmetry at IceCube}},''
  \href{http://dx.doi.org/10.1016/j.physletb.2016.08.066}{Phys. Lett. B
  {\normalfont \bfseries 762} (2016)  116--123},
  \href{http://arxiv.org/abs/1605.09743}{{\normalfont \ttfamily
  arXiv:1605.09743}}.

\bibitem{Babu:2022fje}
K.~S. Babu, P.~S.~B. Dev, and S.~Jana, ``{\em {Probing Neutrino Mass Models
  through Resonances at Neutrino Telescopes}},''
  \href{http://arxiv.org/abs/2202.06975}{{\normalfont \ttfamily
  arXiv:2202.06975}}.

\bibitem{Berezinsky:1975zz}
V.~S. Berezinsky and A.~Y. Smirnov, ``{\em {Cosmic neutrinos of ultra-high
  energies and detection possibility}},''
  \href{http://dx.doi.org/10.1007/BF00643157}{Astrophys. Space Sci.
  {\normalfont \bfseries 32} (1975)  461--482}.

\bibitem{Domokos:1997ve}
G.~Domokos and S.~Kovesi-Domokos, ``{\em {Observation of UHE interactions
  neutrinos from outer space}},'' \href{http://dx.doi.org/10.1063/1.56127}{AIP
  Conf. Proc. {\normalfont \bfseries 433} (1998) no.~1, 390--393},
  \href{http://arxiv.org/abs/hep-ph/9801362}{{\normalfont \ttfamily
  arXiv:hep-ph/9801362}}.

\bibitem{Domokos:1998hz}
G.~Domokos and S.~Kovesi-Domokos, ``{\em {Observation of ultrahigh-energy
  neutrino interactions by orbiting detectors}},''
  \href{http://arxiv.org/abs/hep-ph/9805221}{{\normalfont \ttfamily
  arXiv:hep-ph/9805221}}.

\bibitem{Capelle:1998zz}
K.~S. Capelle, J.~W. Cronin, G.~Parente, and E.~Zas, ``{\em {On the detection
  of ultrahigh-energy neutrinos with the Auger Observatory}},''
  \href{http://dx.doi.org/10.1016/S0927-6505(97)00059-5}{Astropart. Phys.
  {\normalfont \bfseries 8} (1998)  321--328},
  \href{http://arxiv.org/abs/astro-ph/9801313}{{\normalfont \ttfamily
  arXiv:astro-ph/9801313}}.

\bibitem{Fargion:1999se}
D.~Fargion, A.~Aiello, and R.~Conversano, ``{\em {Horizontal tau air showers
  from mountains in deep valley: Traces of UHECR neutrino tau}},'' in {\em
  {26th International Cosmic Ray Conference}}.
\newblock 6, 1999.
\newblock \href{http://arxiv.org/abs/astro-ph/9906450}{{\normalfont \ttfamily
  arXiv:astro-ph/9906450}}.

\bibitem{Fargion:2000iz}
D.~Fargion, ``{\em {Discovering Ultra High Energy Neutrinos by Horizontal and
  Upward tau Air-Showers: Evidences in Terrestrial Gamma Flashes?}},''
  \href{http://dx.doi.org/10.1086/339772}{Astrophys. J. {\normalfont \bfseries
  570} (2002)  909--925},
  \href{http://arxiv.org/abs/astro-ph/0002453}{{\normalfont \ttfamily
  arXiv:astro-ph/0002453}}.

\bibitem{LetessierSelvon:2000kk}
A.~Letessier-Selvon, ``{\em {Establishing the GZK cutoff with ultrahigh-energy
  tau neutrinos}},'' \href{http://dx.doi.org/10.1063/1.1378629}{AIP Conf. Proc.
  {\normalfont \bfseries 566} (2001) no.~1, 157--171},
  \href{http://arxiv.org/abs/astro-ph/0009444}{{\normalfont \ttfamily
  arXiv:astro-ph/0009444}}.

\bibitem{Feng:2001ue}
J.~L. Feng, P.~Fisher, F.~Wilczek, and T.~M. Yu, ``{\em {Observability of earth
  skimming ultrahigh-energy neutrinos}},''
  \href{http://dx.doi.org/10.1103/PhysRevLett.88.161102}{Phys. Rev. Lett.
  {\normalfont \bfseries 88} (2002)  161102},
  \href{http://arxiv.org/abs/hep-ph/0105067}{{\normalfont \ttfamily
  arXiv:hep-ph/0105067}}.

\bibitem{Kusenko:2001gj}
A.~Kusenko and T.~J. Weiler, ``{\em {Neutrino cross-sections at high-energies
  and the future observations of ultrahigh-energy cosmic rays}},''
  \href{http://dx.doi.org/10.1103/PhysRevLett.88.161101}{Phys. Rev. Lett.
  {\normalfont \bfseries 88} (2002)  161101},
  \href{http://arxiv.org/abs/hep-ph/0106071}{{\normalfont \ttfamily
  arXiv:hep-ph/0106071}}.

\bibitem{Bertou:2001vm}
X.~Bertou, P.~Billoir, O.~Deligny, C.~Lachaud, and A.~Letessier-Selvon, ``{\em
  {Tau neutrinos in the Auger Observatory: A New window to UHECR sources}},''
  \href{http://dx.doi.org/10.1016/S0927-6505(01)00147-5}{Astropart. Phys.
  {\normalfont \bfseries 17} (2002)  183--193},
  \href{http://arxiv.org/abs/astro-ph/0104452}{{\normalfont \ttfamily
  arXiv:astro-ph/0104452}}.

\bibitem{Cao:2004sd}
Z.~Cao, M.~A. Huang, P.~Sokolsky, and Y.~Hu, ``{\em {Ultra high energy
  $\nu_\tau$ detection using Cosmic Ray Tau Neutrino Telescope used in
  fluorescence/Cerenkov light detection}},''
  \href{http://dx.doi.org/10.1088/0954-3899/31/7/004}{J. Phys. G {\normalfont
  \bfseries 31} (2005)  571--582},
  \href{http://arxiv.org/abs/astro-ph/0411677}{{\normalfont \ttfamily
  arXiv:astro-ph/0411677}}.

\bibitem{Baret:2011zz}
B.~Baret and V.~Van~Elewyck, ``{\em {High-energy neutrino astronomy: Detection
  methods and first achievements}},''
  \href{http://dx.doi.org/10.1088/0034-4885/74/4/046902}{Rept. Prog. Phys.
  {\normalfont \bfseries 74} (2011)  046902}.

\bibitem{IceCube:2013low}
{\normalfont \bfseries IceCube}, M.~G. Aartsen {\em et al.}, ``{\em {Evidence
  for High-Energy Extraterrestrial Neutrinos at the IceCube Detector}},''
  \href{http://dx.doi.org/10.1126/science.1242856}{Science {\normalfont
  \bfseries 342} (2013)  1242856},
  \href{http://arxiv.org/abs/1311.5238}{{\normalfont \ttfamily
  arXiv:1311.5238}}.

\bibitem{IceCube:2018cha}
{\normalfont \bfseries IceCube}, M.~G. Aartsen {\em et al.}, ``{\em {Neutrino
  emission from the direction of the blazar TXS 0506+056 prior to the
  IceCube-170922A alert}},''
  \href{http://dx.doi.org/10.1126/science.aat2890}{Science {\normalfont
  \bfseries 361} (2018) no.~6398, 147--151},
  \href{http://arxiv.org/abs/1807.08794}{{\normalfont \ttfamily
  arXiv:1807.08794}}.

\bibitem{IceCube:2021rpz}
{\normalfont \bfseries IceCube}, M.~G. Aartsen {\em et al.}, ``{\em {Detection
  of a particle shower at the Glashow resonance with IceCube}},''
  \href{http://dx.doi.org/10.1038/s41586-021-03256-1}{Nature {\normalfont
  \bfseries 591} (2021) no.~7849, 220--224},
  \href{http://arxiv.org/abs/2110.15051}{{\normalfont \ttfamily
  arXiv:2110.15051}}. [Erratum: Nature 592, E11 (2021)].

\bibitem{IceCube:2020abv}
{\normalfont \bfseries IceCube}, R.~Abbasi {\em et al.}, ``{\em {Measurement of
  Astrophysical Tau Neutrinos in IceCube's High-Energy Starting Events}},''
  \href{http://arxiv.org/abs/2011.03561}{{\normalfont \ttfamily
  arXiv:2011.03561}}.

\bibitem{ANTARES:2011hfw}
{\normalfont \bfseries ANTARES}, M.~Ageron {\em et al.}, ``{\em {ANTARES: the
  first undersea neutrino telescope}},''
  \href{http://dx.doi.org/10.1016/j.nima.2011.06.103}{Nucl. Instrum. Meth. A
  {\normalfont \bfseries 656} (2011)  11--38},
  \href{http://arxiv.org/abs/1104.1607}{{\normalfont \ttfamily
  arXiv:1104.1607}}.

\bibitem{Adams:2017fjh}
J.~H. Adams {\em et al.}, ``{\em {White paper on EUSO-SPB2}},''
  \href{http://arxiv.org/abs/1703.04513}{{\normalfont \ttfamily
  arXiv:1703.04513}}.

\bibitem{Eser:2021H6}
J.~Eser, A.~V. Olinto, and L.~Wiencke, ``{\em {Science and mission status of
  EUSO-SPB2}},'' \href{http://dx.doi.org/10.22323/1.395.0404}{PoS {\normalfont
  \bfseries ICRC2021} (2021)  404}.

\bibitem{Cummings:2020ycz}
A.~L. Cummings, R.~Aloisio, and J.~F. Krizmanic, ``{\em {Modeling of the Tau
  and Muon Neutrino-induced Optical Cherenkov Signals from Upward-moving
  Extensive Air Showers}},''
  \href{http://dx.doi.org/10.1103/PhysRevD.103.043017}{Phys. Rev. D
  {\normalfont \bfseries 103} (2021) no.~4, 043017},
  \href{http://arxiv.org/abs/2011.09869}{{\normalfont \ttfamily
  arXiv:2011.09869}}.

\bibitem{Abarr:2020bjd}
{\normalfont \bfseries PUEO}, Q.~Abarr {\em et al.}, ``{\em {The Payload for
  Ultrahigh Energy Observations (PUEO): a white paper}},''
  \href{http://dx.doi.org/10.1088/1748-0221/16/08/P08035}{JINST {\normalfont
  \bfseries 16} (2021) no.~08, P08035},
  \href{http://arxiv.org/abs/2010.02892}{{\normalfont \ttfamily
  arXiv:2010.02892}}.

\bibitem{Vieregg:2021nC}
A.~G. Vieregg, ``{\em {Discovering the Highest Energy Neutrinos with the
  Payload for Ultrahigh Energy Observations (PUEO)}},''
  \href{http://dx.doi.org/10.22323/1.395.1029}{PoS {\normalfont \bfseries
  ICRC2021} (2021)  1029}.

\bibitem{RNO-G:2020rmc}
{\normalfont \bfseries RNO-G}, J.~A. Aguilar {\em et al.}, ``{\em {Design and
  Sensitivity of the Radio Neutrino Observatory in Greenland (RNO-G)}},''
  \href{http://dx.doi.org/10.1088/1748-0221/16/03/P03025}{JINST {\normalfont
  \bfseries 16} (2021) no.~03, P03025},
  \href{http://arxiv.org/abs/2010.12279}{{\normalfont \ttfamily
  arXiv:2010.12279}}.

\bibitem{Aguilar:2021uzt}
J.~A. Aguilar {\em et al.}, ``{\em {Reconstructing the neutrino energy for
  in-ice radio detectors: A study for the Radio Neutrino Observatory Greenland
  (RNO-G)}},'' \href{http://dx.doi.org/10.1140/epjc/s10052-022-10034-4}{Eur.
  Phys. J. C {\normalfont \bfseries 82} (2022) no.~2, 147},
  \href{http://arxiv.org/abs/2107.02604}{{\normalfont \ttfamily
  arXiv:2107.02604}}.

\bibitem{ARA:2019wcf}
{\normalfont \bfseries ARA}, P.~Allison {\em et al.}, ``{\em {Constraints on
  the diffuse flux of ultrahigh energy neutrinos from four years of Askaryan
  Radio Array data in two stations}},''
  \href{http://dx.doi.org/10.1103/PhysRevD.102.043021}{Phys. Rev. D
  {\normalfont \bfseries 102} (2020) no.~4, 043021},
  \href{http://arxiv.org/abs/1912.00987}{{\normalfont \ttfamily
  arXiv:1912.00987}}.

\bibitem{Anker:2020lre}
A.~Anker {\em et al.}, ``{\em {White Paper: ARIANNA-200 high energy neutrino
  telescope}},'' \href{http://arxiv.org/abs/2004.09841}{{\normalfont \ttfamily
  arXiv:2004.09841}}.

\bibitem{deVries:2021BA}
K.~de~Vries {\em et al.}, ``{\em {The Radar Echo Telescope for Neutrinos
  (RET-N)}},'' \href{http://dx.doi.org/10.22323/1.395.1195}{PoS {\normalfont
  \bfseries ICRC2021} (2021)  1195}.

\bibitem{Prohira:2021vvn}
{\normalfont \bfseries Radar Echo Telescope}, S.~Prohira {\em et al.}, ``{\em
  {The Radar Echo Telescope for Cosmic Rays: Pathfinder experiment for a
  next-generation neutrino observatory}},''
  \href{http://dx.doi.org/10.1103/PhysRevD.104.102006}{Phys. Rev. D
  {\normalfont \bfseries 104} (2021) no.~10, 102006},
  \href{http://arxiv.org/abs/2104.00459}{{\normalfont \ttfamily
  arXiv:2104.00459}}.

\bibitem{Prohira:2019glh}
S.~Prohira {\em et al.}, ``{\em {Observation of Radar Echoes From High-Energy
  Particle Cascades}},''
  \href{http://dx.doi.org/10.1103/PhysRevLett.124.091101}{Phys. Rev. Lett.
  {\normalfont \bfseries 124} (2020) no.~9, 091101},
  \href{http://arxiv.org/abs/1910.12830}{{\normalfont \ttfamily
  arXiv:1910.12830}}.

\bibitem{Gandhi:1998ri}
R.~Gandhi, C.~Quigg, M.~H. Reno, and I.~Sarcevic, ``{\em {Neutrino interactions
  at ultrahigh-energies}},''
  \href{http://dx.doi.org/10.1103/PhysRevD.58.093009}{Phys. Rev. D {\normalfont
  \bfseries 58} (1998)  093009},
  \href{http://arxiv.org/abs/hep-ph/9807264}{{\normalfont \ttfamily
  arXiv:hep-ph/9807264}}.

\bibitem{DONUT:2001zvi}
{\normalfont \bfseries DONUT}, R.~Schwienhorst {\em et al.}, ``{\em {A New
  upper limit for the tau - neutrino magnetic moment}},''
  \href{http://dx.doi.org/10.1016/S0370-2693(01)00746-8}{Phys. Lett. B
  {\normalfont \bfseries 513} (2001)  23--29},
  \href{http://arxiv.org/abs/hep-ex/0102026}{{\normalfont \ttfamily
  arXiv:hep-ex/0102026}}.

\bibitem{Coloma:2017ppo}
P.~Coloma, P.~A.~N. Machado, I.~Martinez-Soler, and I.~M. Shoemaker, ``{\em
  {Double-Cascade Events from New Physics in Icecube}},''
  \href{http://dx.doi.org/10.1103/PhysRevLett.119.201804}{Phys. Rev. Lett.
  {\normalfont \bfseries 119} (2017) no.~20, 201804},
  \href{http://arxiv.org/abs/1707.08573}{{\normalfont \ttfamily
  arXiv:1707.08573}}.

\bibitem{Atkinson:2021rnp}
M.~Atkinson, P.~Coloma, I.~Martinez-Soler, N.~Rocco, and I.~M. Shoemaker,
  ``{\em {Heavy Neutrino searches through Double-Bang Events at
  Super-Kamiokande, DUNE, and Hyper-Kamiokande}},''
  \href{http://arxiv.org/abs/2105.09357}{{\normalfont \ttfamily
  arXiv:2105.09357}}.

\bibitem{Alekhin:2015byh}
S.~Alekhin {\em et al.}, ``{\em {A facility to Search for Hidden Particles at
  the CERN SPS: the SHiP physics case}},''
  \href{http://dx.doi.org/10.1088/0034-4885/79/12/124201}{Rept. Prog. Phys.
  {\normalfont \bfseries 79} (2016) no.~12, 124201},
  \href{http://arxiv.org/abs/1504.04855}{{\normalfont \ttfamily
  arXiv:1504.04855}}.

\bibitem{Bustamante:2017xuy}
M.~Bustamante and A.~Connolly, ``{\em {Extracting the Energy-Dependent
  Neutrino-Nucleon Cross Section above 10 TeV Using IceCube Showers}},''
  \href{http://dx.doi.org/10.1103/PhysRevLett.122.041101}{Phys. Rev. Lett.
  {\normalfont \bfseries 122} (2019) no.~4, 041101},
  \href{http://arxiv.org/abs/1711.11043}{{\normalfont \ttfamily
  arXiv:1711.11043}}.

\bibitem{IceCube:2017roe}
{\normalfont \bfseries IceCube}, M.~G. Aartsen {\em et al.}, ``{\em
  {Measurement of the multi-TeV neutrino cross section with IceCube using Earth
  absorption}},'' \href{http://dx.doi.org/10.1038/nature24459}{Nature
  {\normalfont \bfseries 551} (2017)  596--600},
  \href{http://arxiv.org/abs/1711.08119}{{\normalfont \ttfamily
  arXiv:1711.08119}}.

\bibitem{IceCube:2020rnc}
{\normalfont \bfseries IceCube}, R.~Abbasi {\em et al.}, ``{\em {Measurement of
  the high-energy all-flavor neutrino-nucleon cross section with IceCube}},''
  \href{http://arxiv.org/abs/2011.03560}{{\normalfont \ttfamily
  arXiv:2011.03560}}.

\bibitem{Murase:2015ndr}
K.~Murase, {\em {Active Galactic Nuclei as High-Energy Neutrino Sources}},
  \href{http://dx.doi.org/10.1142/9789814759410_0002}{pp.~15--31}.
\newblock 2017.
\newblock \href{http://arxiv.org/abs/1511.01590}{{\normalfont \ttfamily
  arXiv:1511.01590}}.

\bibitem{Bustamante:2020mep}
M.~Bustamante, C.~Rosenstr\o{}m, S.~Shalgar, and I.~Tamborra, ``{\em {Bounds on
  secret neutrino interactions from high-energy astrophysical neutrinos}},''
  \href{http://dx.doi.org/10.1103/PhysRevD.101.123024}{Phys. Rev. D
  {\normalfont \bfseries 101} (2020) no.~12, 123024},
  \href{http://arxiv.org/abs/2001.04994}{{\normalfont \ttfamily
  arXiv:2001.04994}}.

\bibitem{Ng:2014pca}
K.~C.~Y. Ng and J.~F. Beacom, ``{\em {Cosmic neutrino cascades from secret
  neutrino interactions}},''
  \href{http://dx.doi.org/10.1103/PhysRevD.90.065035}{Phys. Rev. D {\normalfont
  \bfseries 90} (2014) no.~6, 065035},
  \href{http://arxiv.org/abs/1404.2288}{{\normalfont \ttfamily
  arXiv:1404.2288}}. [Erratum: Phys.Rev.D 90, 089904 (2014)].

\bibitem{Ioka:2014kca}
K.~Ioka and K.~Murase, ``{\em {IceCube PeV\textendash{}EeV neutrinos and secret
  interactions of neutrinos}},''
  \href{http://dx.doi.org/10.1093/ptep/ptu090}{PTEP {\normalfont \bfseries
  2014} (2014) no.~6, 061E01},
  \href{http://arxiv.org/abs/1404.2279}{{\normalfont \ttfamily
  arXiv:1404.2279}}.

\bibitem{Ibe:2014pja}
M.~Ibe and K.~Kaneta, ``{\em {Cosmic neutrino background absorption line in the
  neutrino spectrum at IceCube}},''
  \href{http://dx.doi.org/10.1103/PhysRevD.90.053011}{Phys. Rev. D {\normalfont
  \bfseries 90} (2014) no.~5, 053011},
  \href{http://arxiv.org/abs/1407.2848}{{\normalfont \ttfamily
  arXiv:1407.2848}}.

\bibitem{Kamada:2015era}
A.~Kamada and H.-B. Yu, ``{\em {Coherent Propagation of PeV Neutrinos and the
  Dip in the Neutrino Spectrum at IceCube}},''
  \href{http://dx.doi.org/10.1103/PhysRevD.92.113004}{Phys. Rev. D {\normalfont
  \bfseries 92} (2015) no.~11, 113004},
  \href{http://arxiv.org/abs/1504.00711}{{\normalfont \ttfamily
  arXiv:1504.00711}}.

\bibitem{Shoemaker:2015qul}
I.~M. Shoemaker and K.~Murase, ``{\em {Probing BSM Neutrino Physics with Flavor
  and Spectral Distortions: Prospects for Future High-Energy Neutrino
  Telescopes}},'' \href{http://dx.doi.org/10.1103/PhysRevD.93.085004}{Phys.
  Rev. D {\normalfont \bfseries 93} (2016) no.~8, 085004},
  \href{http://arxiv.org/abs/1512.07228}{{\normalfont \ttfamily
  arXiv:1512.07228}}.

\bibitem{DiFranzo:2015qea}
A.~DiFranzo and D.~Hooper, ``{\em {Searching for MeV-Scale Gauge Bosons with
  IceCube}},'' \href{http://dx.doi.org/10.1103/PhysRevD.92.095007}{Phys. Rev. D
  {\normalfont \bfseries 92} (2015) no.~9, 095007},
  \href{http://arxiv.org/abs/1507.03015}{{\normalfont \ttfamily
  arXiv:1507.03015}}.

\bibitem{Chew:1952fca}
G.~F. Chew and G.~C. Wick, ``{\em {The Impulse Approximation}},''
  \href{http://dx.doi.org/10.1103/PhysRev.85.636}{Phys. Rev. {\normalfont
  \bfseries 85} (1952) no.~4, 636}.

\bibitem{Magill:2016hgc}
G.~Magill and R.~Plestid, ``{\em {Neutrino Trident Production at the Intensity
  Frontier}},'' \href{http://dx.doi.org/10.1103/PhysRevD.95.073004}{Phys. Rev.
  D {\normalfont \bfseries 95} (2017) no.~7, 073004},
  \href{http://arxiv.org/abs/1612.05642}{{\normalfont \ttfamily
  arXiv:1612.05642}}.

\bibitem{Ge:2017poy}
S.-F. Ge, M.~Lindner, and W.~Rodejohann, ``{\em {Atmospheric Trident Production
  for Probing New Physics}},''
  \href{http://dx.doi.org/10.1016/j.physletb.2017.06.020}{Phys. Lett. B
  {\normalfont \bfseries 772} (2017)  164--168},
  \href{http://arxiv.org/abs/1702.02617}{{\normalfont \ttfamily
  arXiv:1702.02617}}.

\bibitem{Ballett:2018uuc}
P.~Ballett, M.~Hostert, S.~Pascoli, Y.~F. Perez-Gonzalez, Z.~Tabrizi, and
  R.~Zukanovich~Funchal, ``{\em {Neutrino Trident Scattering at Near
  Detectors}},'' \href{http://dx.doi.org/10.1007/JHEP01(2019)119}{JHEP
  {\normalfont \bfseries 01} (2019)  119},
  \href{http://arxiv.org/abs/1807.10973}{{\normalfont \ttfamily
  arXiv:1807.10973}}.

\bibitem{Zhou:2019vxt}
B.~Zhou and J.~F. Beacom, ``{\em {Neutrino-nucleus cross sections for W-boson
  and trident production}},''
  \href{http://dx.doi.org/10.1103/PhysRevD.101.036011}{Phys. Rev. D
  {\normalfont \bfseries 101} (2020) no.~3, 036011},
  \href{http://arxiv.org/abs/1910.08090}{{\normalfont \ttfamily
  arXiv:1910.08090}}.

\bibitem{Drell:1969ca}
S.~D. Drell, D.~J. Levy, and T.-M. Yan, ``{\em {A Field Theoretic Model for
  electron-Nucleon Deep Inelastic Scattering}},''
  \href{http://dx.doi.org/10.1103/PhysRevLett.22.744}{Phys. Rev. Lett.
  {\normalfont \bfseries 22} (1969)  744--748}.

\end{thebibliography}\endgroup

\end{document}